\documentclass[twocolumns]{aa}
\usepackage{graphicx,longtable,lscape,amsmath}
\usepackage[breaklinks,colorlinks,citecolor=blue]{hyperref}
\usepackage[autostyle]{csquotes}
\MakeOuterQuote{"}

\def\gs{{_>\atop^{\sim}}}

\begin{document}

\title{Dynamical complexity in microscale disk-wind systems}

\author{Fabrizio Fiore\inst{1,2}
\and
Massimo Gaspari\inst{3}
\and
Alfredo Luminari\inst{4,5}
\and
Paolo Tozzi\inst{6}
\and
Lucilla de Arcangelis\inst{7}
}

\institute{
INAF - Osservatorio Astronomico di Trieste, via Tiepolo 11, I-34143 Trieste, Italy
\email{fabrizio.fiore@inaf.it}
\and
IFPU - Institute for Fundamental Physics of the Universe, via Beirut 2, I-34151 Trieste, Italy
\and
Department of Astrophysical Sciences, Princeton University, 4 Ivy Lane, Princeton, NJ 08544-1001, USA 
\and
INAF - Istituto di Astrofisica e Planetologia Spaziali, Via del Fosso del Caveliere 100, I-00133 Roma, Italy
\and
INAF - Osservatorio Astronomico di Roma, Via Frascati 33, 00078 Monteporzio, Italy
\and
INAF - Osservatorio Astrofisico di Arcetri, Largo Enrico Fermi, 5, I-50125 Firenze, Italy
\and
Department of Mathematics \& Physics, University of Campania "Luigi Vanvitelli" Viale Lincoln, 5, 81100 Caserta,  Italy
}

\date{February 6th, 2024}

\abstract
{Powerful winds at accretion-disk scales have been observed in the past 20 years in many active galactic nuclei (AGN). These are the so-called ultrafast outflows (UFOs). Outflows are intimately related to mass accretion through the conservation of angular momentum, and they are therefore a key ingredient of most accretion disk models around black holes (BHs). At the same time, nuclear winds and outflows can provide the feedback that regulates the joint BH and galaxy growth.}
{We reconsidered UFO observations in the framework of disk-wind scenarios, both magnetohydrodynamic disk winds and radiatively driven winds. }
{We studied the statistical properties of observed UFOs from the literature and derived the distribution functions of the ratio $\bar \omega$ of the mass-outflow and -inflow rates and the ratio $\lambda_w$ of the mass-outflow and the Eddington accretion rates. We studied the links between $\bar \omega$ and $\lambda_w$ and the Eddington ratio $\lambda={L_{bol}}/{L_{Edd}}$. We derived the typical wind-activity history in our sources by assuming that it can be statistically described by population functions.}
{We find that the distribution functions of $\bar \omega$ and $\lambda_w$ can be described as power laws above some thresholds, suggesting that there may be many wind subevents for each major wind event in each AGN activity cycle, which is a fractal behavior.
We then introduced a simple cellular automaton to investigate how the dynamical properties of an idealized disk-wind system change following the introduction of simple feedback rules. We find that without feedback, the system is overcritical. Conversely, when feedback is present, regardless of whether it is magnetic or radiation driven, the system can be driven toward a self-organized critical state.}
{Our results corroborate the hypothesis that AGN feedback is a necessary key ingredient in disk-wind systems, and following this, in shaping the coevolution of galaxies and supermassive BHs.}

\authorrunning {Fiore et et al.}
\titlerunning {BH disk-wind dynamical systems}

\maketitle

%

\section{Introduction}
The coevolution of black holes (BH) and galaxies is one of the most discussed topics in modern extragalactic astrophysics. BH-galaxy coevolution aims to understand the interplay between active galactic nuclei (AGN), BH accretion, and the evolution of their host galaxies. At the very root of the problem lies the crucial question whether the supermassive BHs (SMBH) systematically found in massive galaxies nuclei are a coincidence or a  necessity. If the growth of a nuclear SMBH is the result of independent processes that produce both BH and galaxy growth, then BHs are a coincidence.  Alternatively, if the two systems influence their respective growth in a nonlinear feedback loop significantly, BHs may be a necessity. 

Today, BHs are an observational evidence in quite a broad mass range, from a few to billions of solar masses (LIGO/VIRGO collaboration 2019; The Event Horizon Telescope collaboration 2019; GRAVITY collaboration 2022). We know that SMBHs in the nuclei of bright central galaxies (BCGs) are the drivers of radio bubbles that heat the intracluster matter (ICM) of cool-core clusters and groups (see McNamara \& Nulsen 2007; Cattaneo et al. 2009; Fabian 2012, Gaspari et al. 2020). This is the main observational evidence so far in favor of the BH necessity scenario. However, BCGs are rare and peculiar galaxies, and their properties cannot be extrapolated straightforwardly to an enormously larger population of galaxies, for which BHs can still be considered freaks of nature.  Indirect hints that the BH necessity scenario can be applied to the population of more common galaxies do exist and include a) the similarity of the star formation and BH accretion histories (Madau \& Dikinson 2014; Aird et al., 2015; Fiore et al., 2017, hereafter F17), b) BH feedback is often invoked as the main driver of the observed correlations between BH and galaxy bulge properties (e.g. Gaspari et al., 2019; Marsden et al., 2020 and references therein), and c) BH feedback is also invoked as the cause of the low growth efficiency of high-mass galaxies (e.g., F17 and references therein). However, similar star formation and BH accretion histories can be the result of the BH coincidence scenario as well, and BH feedback is by no means the only or established main driver of the BH-bulge correlation and star formation quenching. In conclusion, despite very significant observational and theoretical efforts dedicated to the BH-galaxy coevolution problem in the last three decades, the answers to the root questions above remain disturbingly elusive so far, and BHs can still play the awkward role of the "stone-guest at the galaxy evolution supper".

The difficulty of converging toward a solution for the BH-galaxy coevolution problem probably arises because galaxies are complex systems that exchange matter, energy, and entropy with the environment through a network of interactions on many physical and temporal scales. In this respect, they are surprisingly reminiscent of other familiar complex systems: biological cells. This remarkable similarity prompted us to reconsider the problem of BH-galaxy coevolution from a different perspective: from the perspective of complex dynamical systems, which is one of the most successful approaches in evolutionary biology, neurology, and condensed matter. Although our evolving Universe is a clear example of the emergence of complexity from an initially high temperature and energy state, this approach has only been attempted sporadically so far, mainly because that it requires high-quality data and models, both of which have only recently begun to be available. 

In complex dynamical systems, a rich dynamical behavior is generated from the interactions between a large number of subunits. These interactions generate properties that cannot be reduced to the behavior of the single subunit. Or, as in the quote by Phillip Anderson, "more is different". Complex systems share the following main characteristics: first, nonlinearity and sensitivity to initial conditions, the so-called {\it butterfly effect}. Second, Universal emergent properties (e.g., Laughlin 2005) that stem from the interaction of very many degrees of freedom and the naturally developing scale-invariance and power-law correlations. The same statistical theory can describe different systems, regardless of the system details. Third, self-organization is attained through feedback processes, which occur so that the microlevel interactions generate a pattern in the macrolevel that back-reacts onto the subunits. This is called coevolution in evolutionary biology, and it describes how organisms create their environment, which in turn influences the same organisms. This situation intriguingly resembles the BH-galaxy coevolution. Forth, the emergence of a singular behavior is reminiscent of a critical phenomenon, namely a second-order phase transition that is observed in systems that continuously transition between a more ordered and a more disordered phase by the tuning of a single parameter, usually the temperature. This transition occurs at specific values of the thermodynamic variables and exhibits features that are very different from more common first-order transitions that are observed in a more extended region in phase space. At the critical point, the range of spatio-temporal correlations diverges, leading to scale invariance and self-similarity. In recent years, it was observed that macroscopic systems can exhibit critical-like features without a tunable external parameter. In this case, long-range correlations emerge dynamically from the self-organization of local interactions. In this context, a purely Poisson dynamical system can reach a state without a characteristic scale as the critical state when a balance is reached from energy and mass inputs and outputs. It is more interesting when the criticality is self-organized, which is the result of correlated processes that are induced by feedback processes. Self-organized criticality (SOC) is a property found in extremely different physical system, from earthquakes to solar flares, magnetic materials, and the brain (e.g., Bak et al., 1987; Bak 1996; Beggs \& Plenz 2003; de Arcangelis et al., 2006; Friedman et al., 2012; Fontenele et al., 2019). Dynamical systems exhibiting SOC show a highly stochastic behavior that fully reflects their complexity in a phenomenology that escapes any treatment with standard statistical tools.  

We would like to reconsider micro- to macroscale observations and models of BH and galaxy growth by studying them from the perspective of complex dynamical systems. This paper is dedicated to the disk-wind system and therefore deals with microscales from several Schwarzschild radii $r_S$\footnote{$r_S$ is defined as $\frac{2GM_{BH}}{c^2}$, where G is the gravitational constant, $M_{BH}$ is the BH mass and $c$ is the light speed.} (the innermost wind-launch radii) up to the outer edge of the accretion disk and the BH sphere of influence (hundreds of thousand $r_S$). We therefore searched in observations and models for three main characteristics of dynamical systems: 1) An absence of a characteristic scale and criticality, 2) self-organization through feedback processes, and 3) universality. We investigated whether these behaviors might be triggered by BH accretion and winds. Evidence of the same complex dynamics and universality class triggered by BH feedback from micro- to macroscales would highlight the existence of unified principles and point toward the BH necessity scenario. Conversely, if micro- to macroscale scaling relations are mainly produced by Poisson processes, this would point toward a BH coincidence scenario.

In standard geometrically thin optically thick Shakura-Sunyaev (1973) disks, deterministic steady-state solutions exist. The $\alpha$ viscosity that gave the name to these disk models includes both  collisional and turbulent components and is the parameter governing the angular momentum transport. Although turbulence is an intrinsically chaotic process, an $\alpha$ disk retains steady-state solutions because of the key assumption that $\alpha$ is constant or varies slowly at most with respect to the other relevant disk timescales. A natural form of turbulence is produced by magnetic fields (e.g., Balbus 2003), and accretion disks are generally magnetized. Magnetohydrodynamic disk winds (MHDW) thus acquire particular importance in this context. MHDW models are nonlinear. In particular, the Grad-Shafranov equation is highly nonlinear. It expresses the poloidal projection of the momentum balance. The solution of this equation provides the detailed radial structure of the winds, but this is unfortunately an extremely complex equation without a general analytic solution. Simplified solutions based on self-similarity have been found by Blandford \& Payne (1982) and Contopoulos \& Lovelace (1994; also see Koenigl \& Pudritz 2000 and Cui \& Yuan 2019 for reviews). Several 2-dimensional and 3-dimensional numerical simulations of MHDWs have been published so far. In the AGN context, we recall the work of Fukumura et al. (2010) and that by Sadowski et al (2016), Sadowski \& Narayan (2016), Sadowski \& Gaspari (2017; SG17), and Gaspari \& Sadowski (2017; GS17); who simulated a General Relativity MHDW system within a few 100 $r_S$. MHDWs are a natural context for a search for chaotic or complex system behaviors. Several studies have been carried out on theoretical (e.g., Winters, Balbus \& Hawley 2003) and observational grounds (e.g., Sukova, Grzedzielski \& Janiuk 2016). 

Radiation-driven AGN winds (RDW) were discussed in a number of pioneering works during the 1990s (see, e.g., Murray et al. 1995; Proga 2000; Elvis 2000 and references therein) to explain different observations in the framework of AGN atmosphere unification schemes. Hydrodynamical models were developed by various authors (see, e.g., Proga 2007; Kurosawa \& Proga 2009; SG17; and references therein). Radiation driving is at the core of the wind feedback model developed by Silk \& Rees (1998), Fabian (1998), King (2003) and Zubovas \& King (2012, see King \& Pounds 2015 for a review). We discuss both MHDW and RDW in the following sections.

A key process in SMBH feeding is chaotic cold accretion (CCA; Gaspari et al. 2013): The turbulent hot atmosphere recurrently condenses into cold clumps that feed the SMBH from the macro- to the microscale through recurrent/fractal inelastic collisions. These collisions create a self-similar time variability with a flicker-noise 1/f power spectrum (Gaspari et al. 2017). CCA is thus a key example of a chaotic and emergent process in feeding/feedback astrophysics, which is important for understanding some of the results and assumptions in this work. Furthermore, Mineshige et al. (1999) tried to interpret the apparently chaotic X-ray variability of Galactic BH candidates in an SOC framework using a simple cellular automaton to describe the accretion toward the BH.

Winds are a necessary element for dissipating angular momentum and through this, for allowing efficient BH gas accretion. Nuclear (launching radii 10-100 $r_S$) semirelativistic winds are frequently observed in AGN (e.g., Tombesi et al., 2010; 2011; 2012; 2013; Serafinelli et al., 2019; Chartas et al. 2021; see the full reference list in Table \ref{tab:app} in the Appendix). These winds are likely magnetically driven in many cases because a radiation drive can only be efficient for sources that emit at or above their Eddington luminosity (Reynolds 2012). At larger scales (from a fraction of a parsec to a few parsec, i.e., from thousands to tens of thousand $r_S$ ; this is also known as the mesoscale; see GS17), VLT and ALMA interferometric observations have unveiled nuclear dust/molecular disks, rings, and tori in local Seyfert galaxies (Jaffe et al., 2004; Gallimore 2004; Garcia Burillo et al., 2016; Gallimore et al., 2016, Combes et al., 2019; GRAVITY collaboration 2020a; 2020b; 2021). In the few objects in which jets and nuclear disks/rings were observed simultaneously, these structures are not aligned. Because jets are thought to be perpendicular to the inner accretion disks, this suggests that the wide angle relativistic winds emerging from the inner accretion regions can impact the structures around the BH sphere of influence at least in some cases (e.g., tori, rings, or clouds). This can give rise to some feedback between disk winds and gas accretion, which we explore below.

Key open questions still remain unsolved: whether micro- mesoscale feedback is relevant and whether these feedback and nonlinear processes are clues for a complex dynamical system in action. To answer these questions, we first collected observations of semirelativistic winds (or ultrafast outflows, UFOs; Section 2) because they simultaneously are the most common and most energetic nuclear outflows. We then interpret them in the framework of MHDW and radiation driven wind scenarios in Sections 3 and 4. We study the statistical properties of UFOs and derive the distribution functions of the ratio $\bar \omega$ of the mass-outflow and inflow rates in Section 5, where we also study the links between $\bar \omega$ and the Eddington ratio $\lambda=\frac{L_{bol}}{L_{Edd}}$. We then introduce in Section 6 a simple cellular automaton to investigate how the dynamical properties of an idealized disk-wind system change following the introduction of simple feedback rules. We show that if feedback is present, the system can be driven toward an SOC state. We  finally discuss our results in Section 7 and present our conclusions in Section 8.

\section{Sample of observed ultrafast outflows}

\begin{figure}
\includegraphics[width=8.5cm]{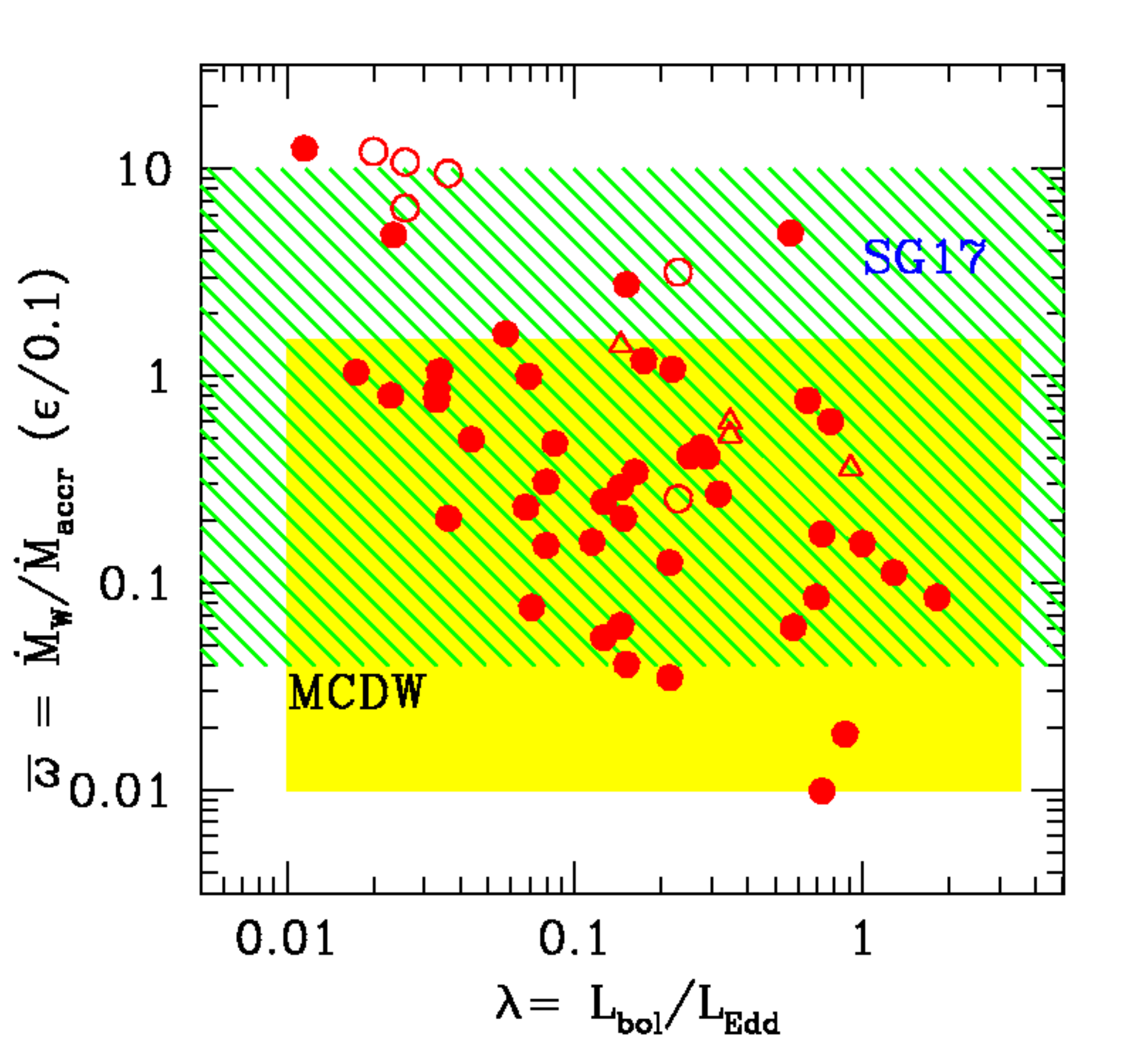}
\caption{$\bar \omega = \frac {\dot M_{\rm w}} {\dot M_{accr}}$ vs. Eddington ratios for a sample of 54 UFO systems in 40 AGN. The filled circles represent z$<0.3$ AGN, the open triangles show z$>0.3$ AGN, and the open circles show lensed AGN. The shaded green area is the expectation of the Sadowski \& Narayan (2016), SG17 and GS17 models, and the yellow area is the expectation of the analytic magneto-centrifugal model (e.g., Cui \& Yuan 2019).}
\label{mdotmufo}
\end{figure}

We compiled UFO observations from the literature that were fit with an homogeneous model (the photoionization model \textrm{XSTAR} within \textrm{XSPEC}). This model provides the gas outflow velocity $v_w$, the column density $N_H$, and the ionization parameter $\xi$. Most of the observations were taken from Tombesi et al 2010, 2011, 2012, 2013 and Gofford et al. 2013, 2015, and they were furthermore incremented with results on single sources. The final sample includes 54 system observations in 40 AGN (see the appendix for details).

To calculate the mass-outflow rate $\dot M_{\rm w}$, we used the approach of Krongold et al (2007), and in particular, their formula in Appendix A2 for the mass-outflow rate of a disk wind in a conical geometry,

\begin{equation}
\dot M_w=0.8 \pi m_p f N_H v_w r  = 6.3055 \times 10^{-24} N_H v_w r M_{\odot}/yr,
\end{equation}

\noindent
where $f$ is a function of the angle of the line of sight to the central source and the accretion disk plane and the angle formed by the wind with the accretion disk. $r$ is the distance of the illuminated face of the wind from the central luminosity source. The numerical estimate on the right-hand side applies when $f=1.5$, corresponding to a roughly vertical disk wind and an average line-of-sight angle of $30^{\circ}$.

To account for the scaling of the mass-outflow rate with source activity, we studied the mass-outflow rate normalized by the mass-accretion rate $\bar \omega=\frac{\dot M_{\rm w}}{\dot M_{accr}}$, where $\dot M_{accr}=\frac{L_{bol}} {\epsilon c^2}$, and $\epsilon$ is the radiative efficiency. We show in the next section that $\bar \omega$ has a profound significance in the context of accretion models.

While $N_H$ and $v_w$ in eq. 1 are results of fitting the data with a photoionization model, $r$ cannot be directly estimated from the X-ray data in most cases. It is usually assumed in the literature either as the radius at which the observed velocity equals the escape velocity, $r_{min}=\frac{2GM_{BH}}{v_w^2}$, or from the definition of the ionization parameter $\xi= \frac{L_{ion}}{{n_e r^2}}$, where $L_{ion}$ is the luminosity above the hydrogen ionization threshold. It follows from this that $r_{ion}\approx \frac{L_{ion}}{1.23 \xi N_H} \frac{\Delta r}{r}$ because $n_e\approx 1.23 n_H$ for a fully ionized gas with solar abundances, and $N_H=n_H \Delta r$, where the thickness of the absorber is $\Delta r<<r$. If $r=r_{min}$,
\begin{equation}
   \dot M_{\rm w}(r
   _{min}) = 0.88 \frac{N_H} {10^{24}} \frac {M_{BH}} {10^8} \left(\frac{v_w/c} {0.1} \right)^{-1} M_\odot/yr,
\end{equation}
\noindent
and
\begin{equation}
\begin{split}
   \bar \omega (r_{min})=  5.03 \frac{N_H} {10^{24}} \frac {M_{BH}} {10^8} \left(\frac{v_w/c} {0.1}\right)^{-1} \Big( \frac {L_{bol}} {10^{45}} \Big)^{-1} \frac {\epsilon} {0.1}  \\
   = 4.53  \frac{N_H} {10^{24}} \left(\frac{v_w/c} {0.1}\right)^{-1} \left(\frac {\lambda} {0.1}\right)^{-1} \frac {\epsilon} {0.1} .
\end{split}
\end{equation}

\noindent
In this formulation, $\dot M_{\rm w}(r_{min})$ depends linearly on $N_H$ and $M_{BH}$ and inversely on $v_w$, while $\bar \omega(r_{min})$ depends linearly on $N_H$ and inversely on $v_w$ and $\lambda$. The range covered by the measured $N_H$ is between $10^{22}$cm$^{-2}$ and $10^{24}$cm$^{-2}$, that is, 2 dex, with a typical uncertainty of about 0.2-0.5 dex. The range covered by $v_w$ is 0.05-0.5c, about 1 dex,  with a typical uncertainty of 0.1-0.2 dex. The range of $\lambda=\frac {L_{Bol}} {L_{Edd}}$ spans $\gs 2$dex (see the appendix for details). The uncertainty associated with the unknown geometry of the wind is about 0.2-0.3 dex (Krongold et al. 2007). While the relative statistical and systematic uncertainties on $\dot M_{\rm w}(r_{min})$ and $\bar \omega(r_{min})$ are generally small, $\dot M_{\rm w}(r_{min})$ can by construction be regarded as a lower limit to the real mass-outflow rate.

If $r=r_{ion}$,
\begin{equation}
   \dot M_{\rm w}(r_{ion})=0.24 \frac{L_{ion}} {10^{44}} \frac {10^4} {\xi} \frac{v_w/c} {0.1} \frac {\Delta r/r} {0.1} M_\odot/yr
\end{equation},
\noindent
and
\begin{equation}
   \bar \omega(r_{ion})= 1.37 \frac {10^4} {\xi} \frac{v_w/c} {0.1} \frac {\epsilon} {0.1} \frac {\Delta r/r} {0.1} 
\end{equation},
\noindent
assuming that $L_{ion}\approx 0.1 L_{bol}$ (see, e.g., Lusso et al. 2012). In this formulation, $\dot M_{\rm w}(r_{ion})$ depends linearly on both $L_{ion}$ and $v_w$ and inversely on $\xi$. It does not depend on $N_H$. $\bar \omega(r_{ion})$ depends linearly on $v_w$ and inversely on $\xi$. The range covered by $\xi$ is about 3 dex, and its uncertainty is about 0.1-0.2 dex. $\Delta r/r$ is generally difficult to estimate from the data. The few attempts have been made for warm absorbers rather than UFOs (see, e.g., Krongold et al. 2007). $\Delta r/r$ is thought to be about 0.1 or lower in conical outflows. Critically, different flows can have quite different $\Delta r/r$, and therefore, the scatter in $\Delta r/r$ can dominate the scatter on $\bar \omega(r_{ion})$. Therefore, assuming a fixed value may artificially reduce the intrinsic scatter of $\bar \omega(r_{ion})$ in a sample of flows. In the end, the real normalized mass-outflow rate should be somewhere in between $\bar \omega(r_{min})$ and $\bar \omega(r_{ion})$, and the scatter on $\bar \omega(r_{min})$ in a sample of flows is probably a lower limit to the real scatter. In our sample of 54 flows, the median values of $r_{min}$ and $r_{ion}$ are about $40r_S$ and $300\frac {\Delta r/r} {0.1} r_S$, and the average of the difference between $\bar \omega(r_{min})$ and $\bar \omega(r_{ion})$ is 0.8dex. We plot in Figure \ref{mdotmufo} the average between these two extremes, computed as $log(\bar \omega) = <log(\bar \omega(r_{ion}))-log(\bar \omega(r_{min}))> = log(\bar \omega(r_{min})) + 0.4$, as a function of the AGN luminosity in Eddington units $\lambda$. $\dot M_{accr}$ was estimated from the bolometric luminosity as $\dot M_{accr}=\frac {L_{bol}} {c^2 \epsilon}$ assuming a radiative efficiency $\epsilon=0.1$. The wind mass-outflow rates in Figure \ref{mdotmufo} were corrected for relativistic effects following Luminari et al. (2020). The corrections were usually small, however, and the results obtained with the uncorrected data are quite similar to the results presented below.

Two features stand out in Figure \ref{mdotmufo}. The first feature is the expected correlation between $\bar \omega$ and $\lambda$ (see eq. 6). The Spearman-rank correlation coefficient of the $\bar \omega-\lambda$ correlation is -0.45 for 52 degrees of freedom. The best-fit slope of the logarithmic $\bar \omega-\lambda$ correlation in Figure \ref{mdotmufo} is $0.7\pm0.2$.  Since $\bar \omega=\frac{\dot M_{\rm w}}{\dot M_{accr}}$ and $\lambda=\frac{L_{bol}}{L_{Edd}}=\frac{\dot M_{accr}} {\dot M_{Edd}}$, then
\begin{equation}
  \bar \omega = \frac{1} {\lambda} \times \frac{\dot M_{\rm w}} {\dot M_{Edd}} = \frac{\lambda_w} {\lambda},
\end{equation}
\noindent
that is, $\bar \omega$ scales with the ratio of the wind Eddington rate and the Eddington rate. The second feature is that the observed $\bar \omega$ spans more than 3 dex (more than 2 dex at all $\lambda$). This range is wider than the statistical or systematic uncertainty on $\dot M_{\rm w}$, and it is also robust against the presence of outliers. Some of the systems with the highest $\bar \omega$ in Figure \ref{mdotmufo} are lensed systems, in which an usually large uncertainty on the lens magnification adds to the other systematic and statistical errors. Other high $\bar \omega$ systems are low-luminosity highly obscured Seyfert 2 galaxies, where the photoelectric cutoff extends to the energies of the UFO absorption features, which means that they are more complex to characterize. Even when these problematic sources are excluded, the picture remains substantially unchanged. As discussed above, the scatter in Figure \ref{mdotmufo} is most likely a lower limit to the real scatter. In the following section, we compare these findings with the prediction of disk-wind models. 

\section{Magnetohydrodynamic disk-wind scenario}

Nuclear winds are intimately related to mass accretion through the conservation of angular momentum, and they are likely magnetically driven in most cases (e.g., Reynolds 2012). More recently, Wang et al. (2022) performed MHD simulations of winds from thin accretion disks. The authors suggested that events resembling observed UFOs can certainly be generated within about 600$r_S$. Beyond this radius, winds have a smaller ionization parameter than observed UFOs. In our sample of 54 UFOs, all but two have $r_{min}<600 r_S$ and 60\% have $r_{ion}<600 r_S$. We also note that the ionization of the wind is governed by how radiative feedback is produced/injected in the wind (in addition to $\Delta r/r$). The source geometry of the ionization radiation is definitely poorly constrained. It can be an extended corona (which produces high ionization at relatively large radii) or a lamp-post region that is confined in the innermost region of the accretion disk (which mainly produces ionization at small radii). 

The MHDWs are therefore a plausible way to describe the observed UFOs. We do not discuss the fate of these winds and whether they will produce energy-driven or momentum-driven outflows in the galaxy. 

In MHDW systems, the conservation of momentum rate reads (see the reviews of Konigl \& Pudritz 2000 and Pudritz et al. 2007)
\begin{equation}
    \dot J_d=\dot M_{in} r_0^2 \Omega_0 = \dot J_w=\dot M_{out} \Omega_0 r_A^2,
\end{equation}
\noindent
where $\dot J_d$ and $\dot J_w$ are the disk and wind momentum rates, respectively, $\dot M_{in}$ and $\dot M_{out}$ are the mass-accretion and -outflow rates, $r_0$ and $r_A$ are the wind launch radius and the Alfv\'en radius, and $\Omega_0$ is the Keplerian angular velocity at the launch radius (every annulus of the disk rotates close to its breakup speed). Mass accretion and mass outflows are thus intimately connected. No accretion is possible without the launch of massive winds from the system, and the Alfv\'en radius is the lever arm of the outflow. Because $\frac{r_A}{r_0}>1$, relatively small mass outflows can cause relatively large inflows: The removal of angular momentum by the winds allows matter to accrete inward. The Alfv\'en radius marks the extent of the region in which magnetic stresses dominate the flow. this follows the poloidal magnetic field lines anchored to the disk. Therefore, beyond this radius, the outflow terminates and corotates with the disk, reaching a terminal speed $v_w = v_{\infty} = \sqrt{2} v_0 \frac{r_A}{r_0}$, where $v_0$ is the wind launching velocity at $r_0$. The acceleration from the disk occurs through a centrifugal effect, and therefore, these models are called magneto-centrifugal disk wind (MCDW). 

In all disk-wind models, the accretion and outflow rates $\dot M_{in}$ and $\dot M_{out}$ depend on the radius. The outflow rate probed by an X-ray UFO observation probably concerns the phase when the wind emerges from the system at tens or hundreds of Schwarzschild radii $r_S$, well before the entrainment phase, in which the outflow rate can be significantly loaded (see, e.g., Serafinelli et al. 2019), and therefore, it is quite justifyable to assume $\dot M_{w}=\dot M_{out}$. The observed X-ray and bolometric luminosities are produced in regions from a few to some dozen $r_S$, and therefore, we identify $\dot M_{in}$ with $\dot M_{accr}$. Therefore, $\frac{\dot M_{out}}{\dot M_{in}} \approx \bar \omega$. In MCDW models, $\bar \omega = (\frac{r_0}{r_A})^2 \in[0.01,1]$ for typical $\frac{r_0}{r_A}$ ratios, and $r_A \gs r_0$. The yellow area in Figure \ref{mdotmufo} marks the expectations of the MCDW model. $\bar \omega$ depends on the intensity of the poloidal magnetic field at the radius $r_0$: The energy density of the magnetic field should equal the kinetic energy density in the wind, or
\begin{equation}
v_A  = v_p = \frac{B_p}{(4\pi \rho)^{0.5}} , 
\end{equation}
\noindent
where $V_A$ is the poloidal Alfv\'en velocity, $v_p$ is the poloidal velocity , $B_p$ is the poloidal magnetic field, and $\rho$ is the wind density. From eq. 8, it follows that
\begin{equation}
r_A = \frac{B_p}{\Omega_0 (4\pi \rho)^{0.5}}.  
\end{equation}
\noindent
While it is useful to emphasize the key role of winds in accretion systems in a direct and simple way, the MCDW models described above are highly idealized and are based on a simplified analysis. They do not include important physics such as thermal and radiation pressure and general relativity effects. Sadowski et al. (2016), Sadowski \& Narayan (2016), SG17, and GS17 developed more realistic MHDGR models and simulated the accretion/outflows in a box of $\sim100 r_S$ for systems with a wide range of $\lambda$. They found a generally high kinetic efficiency $\epsilon_w = \frac{L_w}{\dot M_{\bullet} c^2} \approx 0.02-0.04$ that was roughly constant with $\lambda$ from highly sub-Eddington to super-Eddington regimes. Here, $\dot M_{\bullet}$ is the accretion rate through the event horizon, and it thus contributes to the BH growth. Due to GR effects, $\dot M_{\bullet}$ can be lower than $\dot M_{\rm accr}$ (the accretion rate at larger radii, where most of the radiative luminosity is produced), and thus, $\epsilon_w=\frac{L_w}{\dot M_{accr} c^2} \approx 0.005-0.01$. Since $L_w=1/2 \dot M_{\rm w} v_w^2=\epsilon_w \dot M_{\bullet} c^2$, then $\bar \omega=\frac{\dot M_{\rm w}}{\dot M_{accr}} \sim 200 (\frac{v_w/c}{0.1})^{-2} \times \epsilon_w$. The shaded green area in Figure \ref{mdotmufo} marks the expectation of this model for $v_w/c \in[0.05,0.5]$, the velocity range of most observed UFOs. Figure \ref{mdotmufo} shows that all observed UFO $\bar \omega$s are consistent with the predictions of the MHDGR model considering the statistical and systematic uncertainties. 

\section{Radiation-driven disk-wind scenario}

Winds can be efficiently driven by radiation when the luminosity of the central source is sufficiently high and when the gas is sufficiently optically thick, as illustrated by the Eddington equation,
\begin{equation}
    L_{edd}=\frac {4\pi GM_{BH}m_p c} {\sigma},
\end{equation}
where $M_{BH}$ is the black hole mass, $m_p$ is the mass of the proton, and $\sigma$ is an absorption cross section. For completely ionized gas and no dust, $\sigma=\sigma_T$, where $\sigma_T$ is the Thompson cross section, while when the gas is not completely ionized and/or dusty $\sigma>>\sigma_T$, because the line and dust opacities can be orders of magnitude higher than the Thomson-scattering opacity. However, in the innermost nuclear regions, below tens or hundreds of gravitational radii, the gas is likely fully ionized and dust free, and therefore, an efficient wind-driving through radiation pressure requires huge luminosities and dense gas clouds. When the gas is sufficiently dense (Compton thick, $N_H\gs10^{24}$ cm$^{-2}$), each photon would scatter at least once before it escapes, yielding all its moment. Therefore, $\dot M_w v_w \approx L_{bol}/c = L_{Edd}/c \times \lambda$, and $\dot M_w=L_{bol}/(v_w c) = L_{edd}/(v_w c) \times \lambda$. If $\dot M_{acc}=L_{bol}/\eta c^2$, then $\bar \omega = \dot M_w / \dot M_{accr} = \eta c/v_w$.  

The identification of UFOs with these continuous winds  is problematic, however, because Compton-thick intervening gas would strongly depress the nuclear X-ray flux, making the detection of absorption lines difficult if not impossible. A solution for this apparent conundrum is that UFOs are episodic, as in the scenario suggested by King \& Pounds (2015). The wind is launched when the central engine runs at $L_{bol}\sim L_{Edd}$. When this phase is short, the wind is not continuous, a shell moves away from the nucleus, and by expanding, it gradually decreases its column density. At some point, absorption lines become observable. In this scenario, the wind mass-outflow rate at the launch radius is $\dot M_w\approx L_{Edd}/(v_w c)$. If $N_H\propto 1/r_{in}$ (King \& Pounds 2015). Then, $\dot M_w$ should be nearly constant with distance from the nucleus and time because $\dot M_w \propto N_H \times r \times v_w$ (see eq. 1). Finally, the scatter of the observed $\bar \omega$ in this scenario depends on $\eta$ and $v_w/c$, as in the case of continuous winds. 

The fast wind expands until it collides with the ambient gas, forming a reverse shock that slows the nuclear wind down, and a forward shock expands in the interstellar gas of the BH host galaxy. The AGN radiation can efficiently cool the shocked gas, deciding the nature and the fate of the shock. When the cooling is strong, the shocked gas looses most of its energy and only transmits its ram pressure $\dot M v_w$ to the post-shock gas (momentum-driven flows). Conversely, when cooling is negligible, the flow maintains most of its energy and expands adiabatically in the interstellar gas (energy-driven flows). Since cooling is mostly driven by the AGN radiation field, the cooling time near the BH will be shorter than the flow time and the flow will be driven by momentum. The cooling time increases as $R^2$, but the flow timescale increases linearly with distance. This means that beyond a critical distance, the cooling time can exceed the flow time, and the flow becomes energy driven (Zubovas \& King 2012; King \& Pounds 2015). UFOs probe nuclear flows, and they therefore most likely belong to the momentum-driven regime. It is thus expected that most of the sources in compilations of UFOs and galaxy-scale outflows are compatible with momentum-driven flows (e.g., Bonanomi et al. 2023). Because nuclear wind is likely short lived, the observed UFOs may well not be causally related with the more long-lived galaxy-scale winds.

\section{Population properties versus single-system evolution}

\begin{figure*}
\begin{tabular}{cc}
\includegraphics[width=8.5cm]{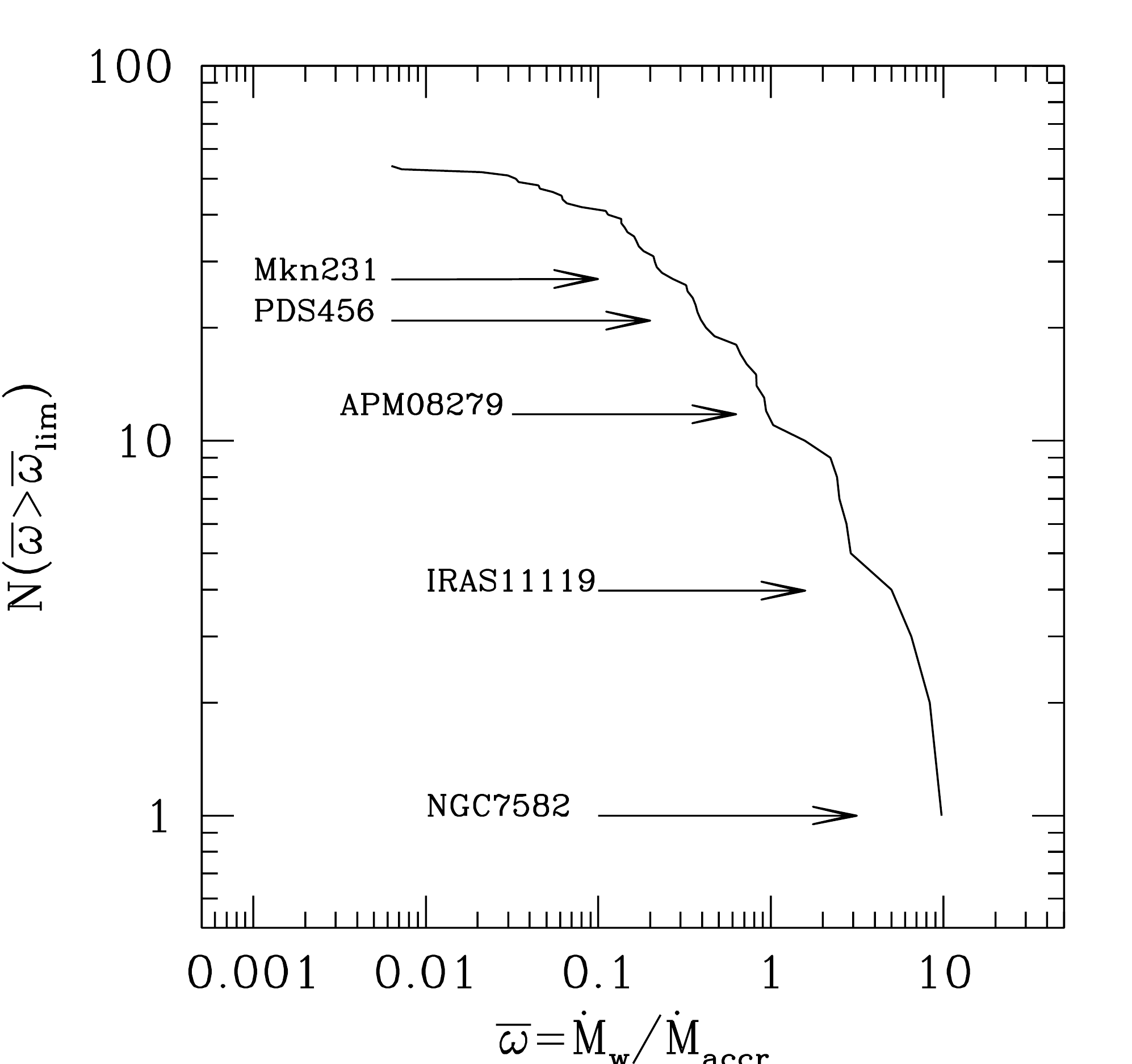}
\includegraphics[width=8.5cm]{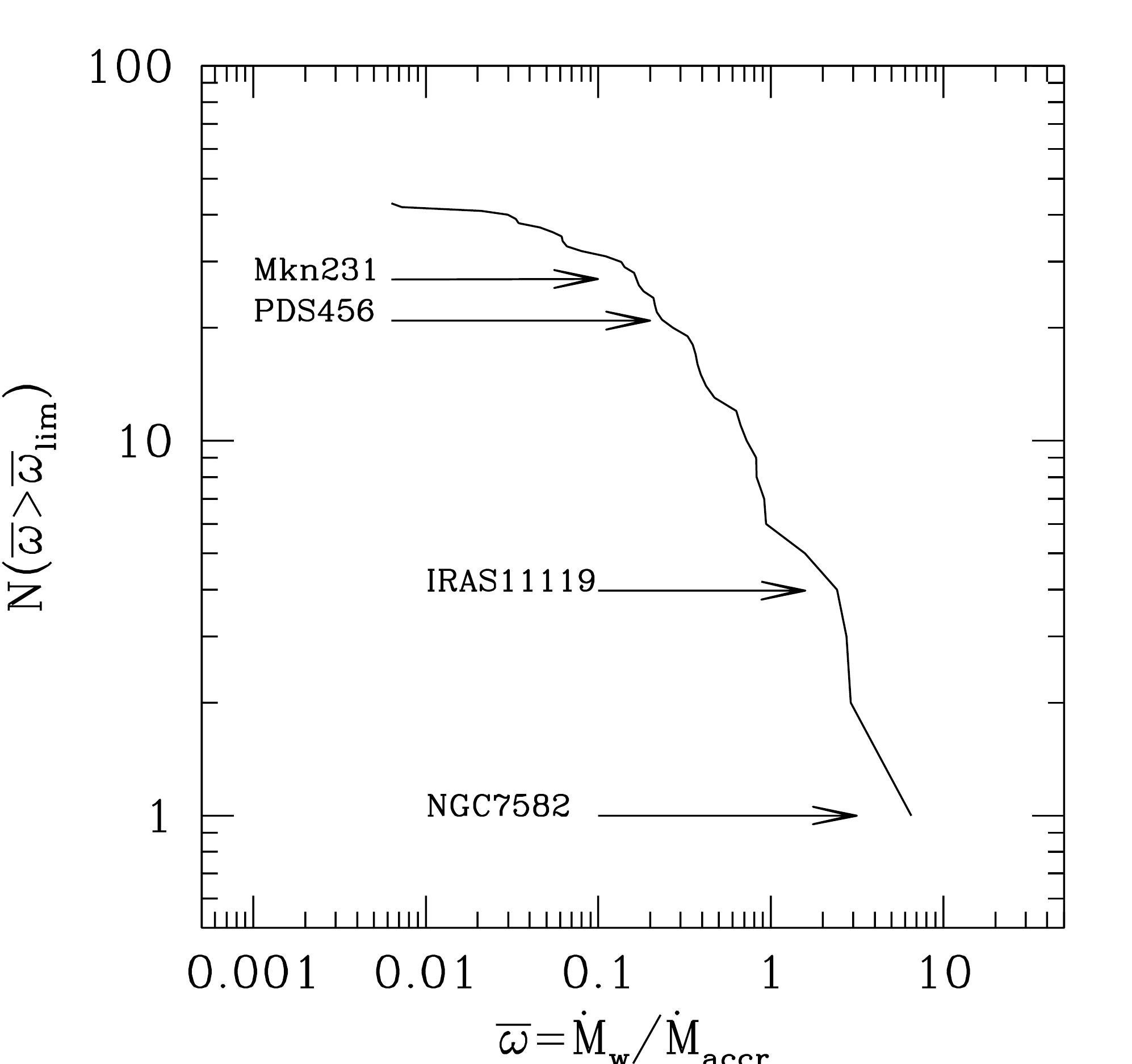}
\end{tabular}
\caption{Cumulative distribution function of the number of systems in which $\bar \omega$ exceeds a limit value. [Left panel]: 54 individual systems in 40 AGN. [Right panel]: 43 individual system in 34 sources at z$<$0.3}
\label{omegacounts}
\end{figure*}

The main feature highlighted by Figure \ref{mdotmufo} is the broad range of $\bar \omega$ that is covered by existing observations. The range is broader than the statistical errors on single determinations and the systematic errors that are certainly present in a biased and inhomogeneous sample such as we used in this analysis. Therefore, the wide $\bar \omega$ range is likely intrinsic to the wind systems, suggesting that a broad distribution of this key parameter is produced by physical disk-wind dynamic systems. As discussed in the previous sections, a broad range of $\bar \omega$ can be easily explained by either the MHDW or the RDW scenarions, in which different physical drivers cause the scatter. To further investigate this crucial point, we calculated the cumulative and differential distribution functions of the number of systems in which $\bar \omega$ exceeds a limit value as a function of this limit value. These functions are the equivalent of the classic LogN-LogS functions in astrophysics. The distribution functions are plotted in Figure \ref{omegacounts} for the full sample (left panel) and for the sources at z$<0.3$, that is, high-redshift luminous quasars and lensed sources are excluded. Luminous quasars can be powered by accretion flows quite different than low-redshift AGN, while the uncertainty on the wind and accretion parameters in the lensed sources are increased because it is difficult to model the amplification at different spatial scales. From now on, we consider the sample of AGN at z$<0.3$, which is both more homogeneous and robust. 

Below $\bar \omega\sim 0.05$, the cumulative distribution function in the right panel of Figure \ref{omegacounts} flattens considerably, most probably because selection effects limit the completeness of the UFO sample. Therefore, to be representative of the global AGN population, the functions in Fig. \ref{omegacounts} must be corrected for these observational selection effects. Winds with a low mass-outflow rate can only be detected in spectra with the highest signal-to-noise ratio, which are rarer. This produces a flattening of the functions toward low $\bar \omega$ values. A correction for this selection effect would increase the number of systems with small $\bar \omega$. We evaluated a correction for incompleteness as described in the appendix. Figure \ref{omegacountsc} shows the normalized cumulative and differential (per decade) distribution functions of $\bar \omega$ for the sample of UFO systems at $z<0.3$, corrected for incompleteness. The distribution functions were normalized by dividing by the total number of sources (34) in which 43 independent UFOs were observed.

\begin{figure*}
\begin{tabular}{cc}
\includegraphics[width=8.5cm]{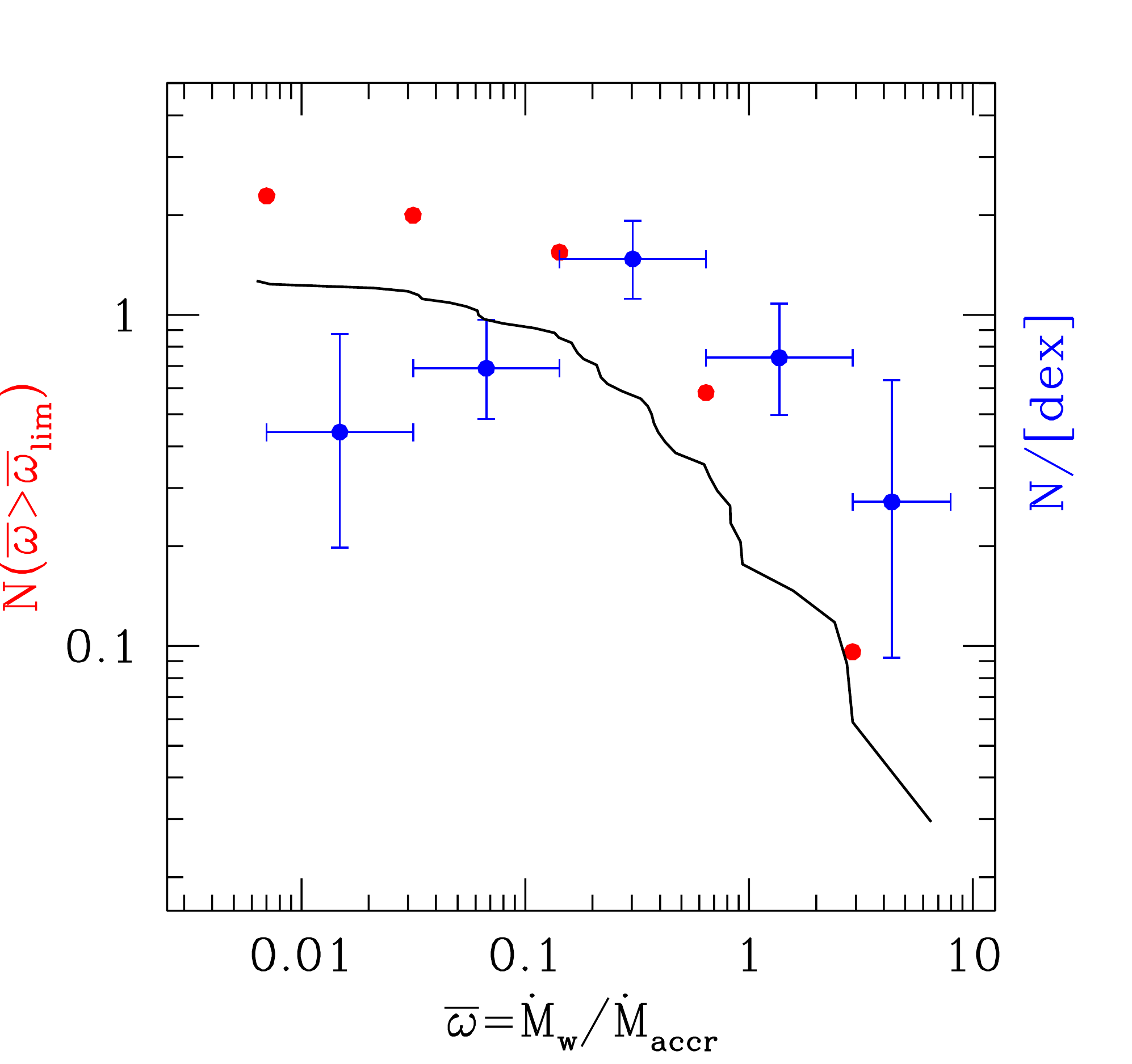}
\includegraphics[width=8.5cm]{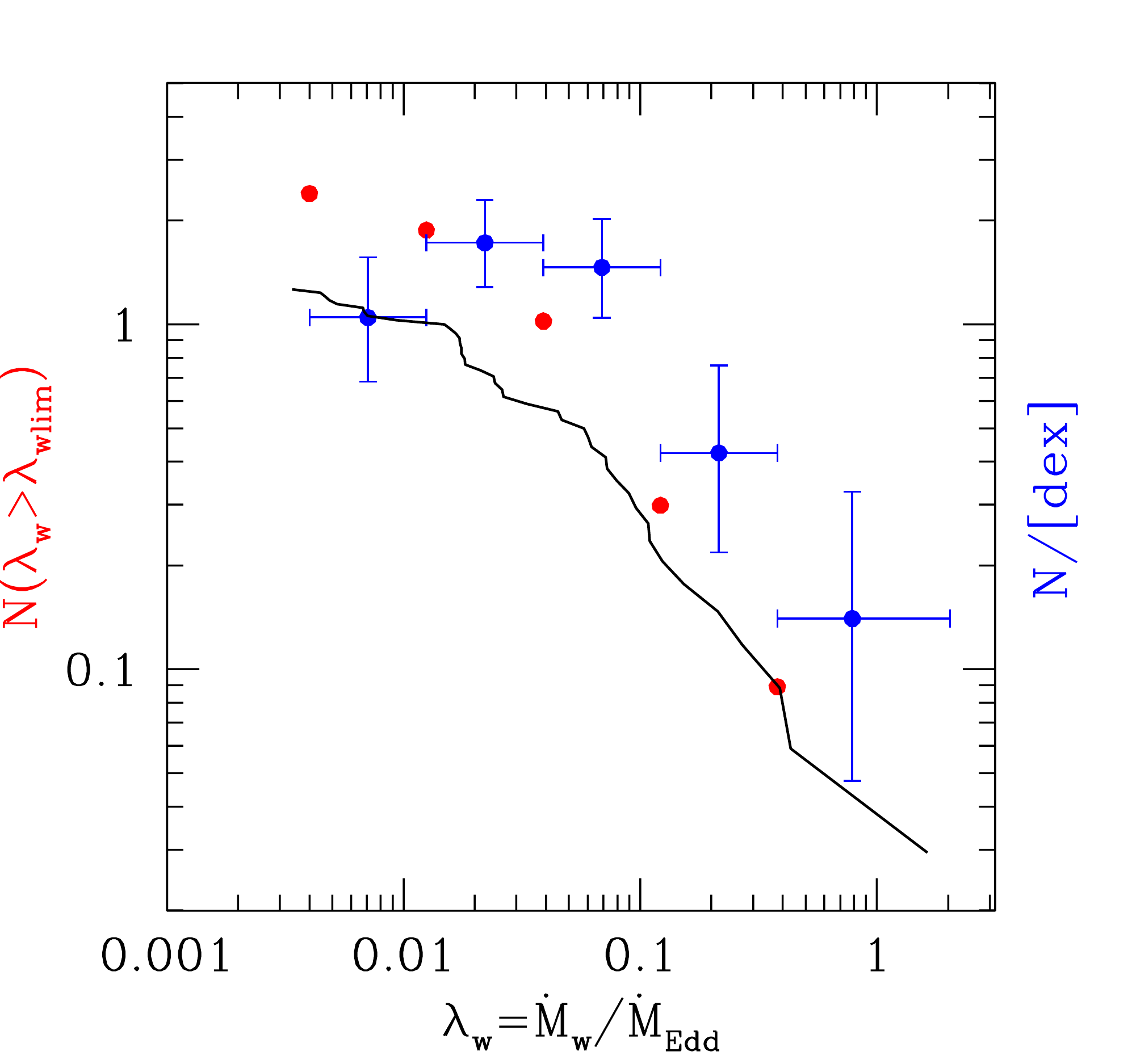}
\end{tabular}
\caption{Normalized, cumulative (red), and differential (per decade; blue) binned distribution functions of $\bar \omega=\frac{\dot M_{\rm w}}{\dot M_{accr}}$ [left panel] and $\lambda_w=\frac{\dot M_{\rm w}} {\dot M_{Edd}}$ [right panel] for 43 individual UFO systems in 34 AGN at z$<$0.3 (both corrected for selection effects; see the appendix). The black curve shows the uncorrected $\bar \omega$ and $\lambda_w$ cumulative distribution functions. The black curve in the left panel is the same as the black curve in the right panel of Figure \ref{omegacounts} .}
\label{omegacountsc}
\end{figure*}

The power-law slopes of the $\bar \omega$ cumulative and differential (per decade) distribution functions above $\bar \omega=0.2$ are $-0.86^{-0.40}_{+0.27}$ and $-0.57^{-0.39}_{+0.29}$, respectively (90\% confidence error). The power-law slopes of the $\lambda_w$ cumulative and differential distribution functions above $\lambda_w=0.1$ are $-1.04^{-0.61}_{+0.42}$ and $-1.00^{-0.76}_{+0.44}$, respectively. In the $\bar \omega$ and $\lambda_w$ ranges 0.007-0.14 and 0.004-0.04, the differential distributions are flat, with a slope consistent with 0. The application in the next years of this type of analysis to less biased and complete AGN samples, such as the \textrm{SUBWAYS} sample (Mehdipour et al., 2022; Matzeu et al, 2022) 
is crucial to reduce the statistical and systematic uncertainties.

\begin{figure}
\includegraphics[width=8.5cm]{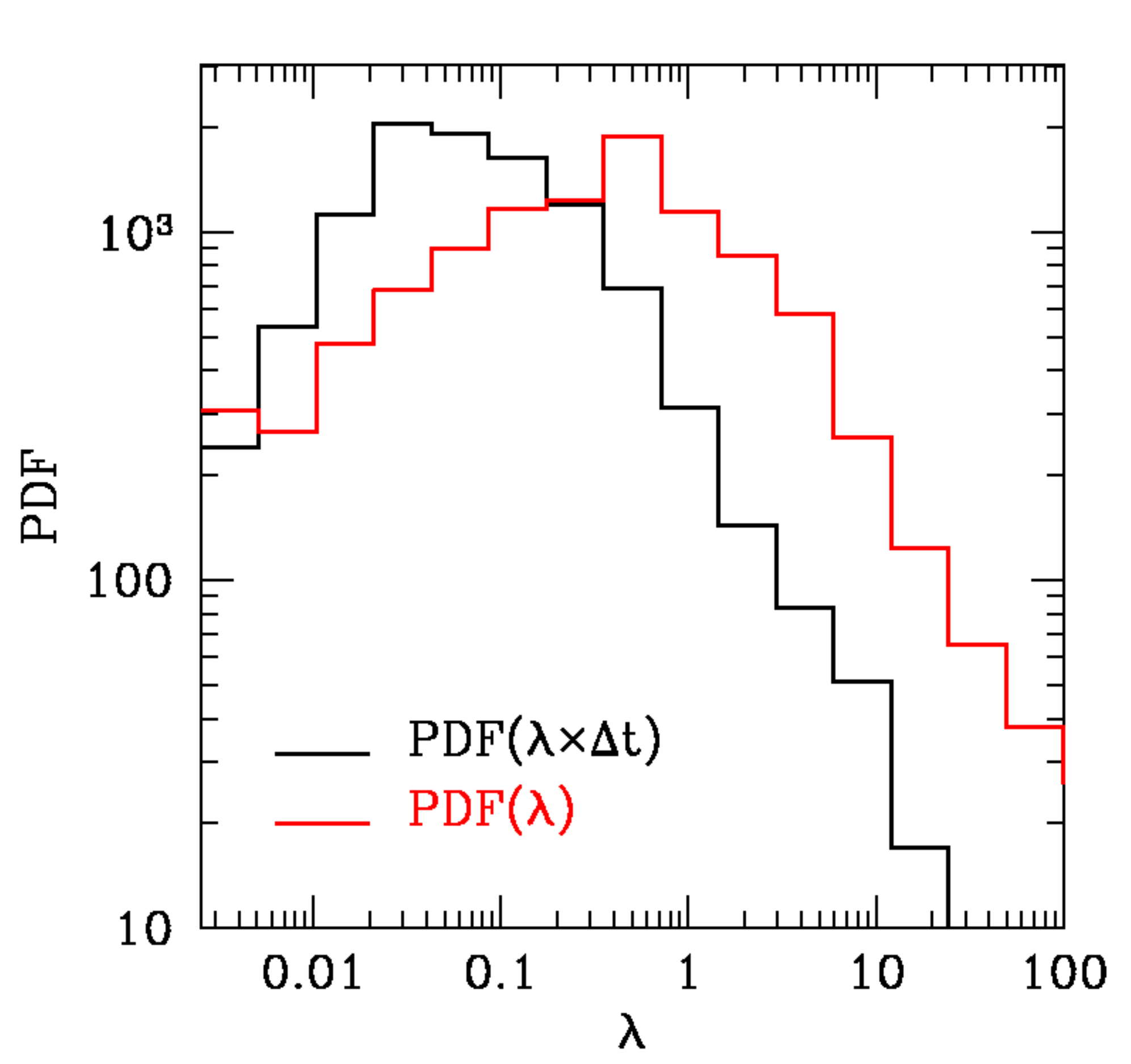}
\caption{Expected probability distribution function of $\lambda$ (red histogram) and $\lambda \times \Delta t$ based on the observed distributions of $\bar \omega$ and $\lambda_w$ in a sample of UFO at z$<$0.3.}
\label{lamdistr}
\end{figure}

When the wind-activity timescale is shorter than the AGN timescale, as is likely in many cases (e.g., Nardini \& Zubovas 2018), and when the population functions in Figure \ref{omegacountsc} are drawn from independent snapshots, we can argue that the wind-activity history in each single source can be statistically described by these population functions. If this is the case, the message suggested by the analysis in the previous sections is that UFOs can be seen as series of sporadically launched quasi-spherical shells, with several or many relatively small wind events for each main wind event in each AGN activity cycle. This scenario is consistent with both MHDW and RDW. Expanding on this idea, we suggest that wind events may have a fractal behavior, analogous to what was suggested by Gaspari et al.~(2017) for CCA. Since 
\begin{equation}
\lambda=\frac{\dot M_{accr}}{\dot M_{Edd}} = \frac{\lambda_w}{\bar \omega},
\end{equation}
\noindent
$\lambda$ should follow a power-law distribution down to an inflection point as well. $\lambda_w$ and $\bar \omega$ are both independent of the BH mass. Thus, the distribution of $\lambda$ can be obtained from the distribution functions of $\bar \omega$ and $\lambda_w$, both corrected for selection effects. We therefore assumed broken power-law distributions for $\bar \omega$ and $\lambda_w$ (with the parameters given above), and randomly generated 10,000 $\bar \omega$ and $\lambda_w$ values following these distributions, mimicking 10,000 wind and accretion events in the active lifetime of a single source. The $\lambda$ distribution function obtained in this way is plotted in Figure \ref{lamdistr} (red histogram) and resembles a log-normal distribution that peaks at $\lambda\approx 0.5$. To compare this distribution function with the observed $\lambda$ distribution functions from AGN surveys, again assuming that the population properties reflect the evolution of single systems, we need to assess the time spent by the system on each $\lambda$ value. To estimate this timescale, we considered that the relativistic winds carry momentum and energy and that they are likely at a wide angle (e.g., Nardini et al. 2015; Luminari et al. 2018). The wind propagates in the direction of least contrast, that is, nearly perpendicularly to the accretion disk, until it encounters matter with a density that is high enough to shock (e.g., Menci et al. 2019 and references therein). This is likely to occur at the limit of the BH sphere of influence. Beyond this region, the geometry of the matter is not governed by the BH gravity and is more complex, and it is in many case not aligned with the accretion disk. VLT interferometric observations and ALMA observations have unveiled nuclear dust/molecular disks/rings/tori on scales down to a fraction of one parsec in local Seyfert galaxies (Jaffe et al., 2004; Garcia Burillo et al., 2016; Gallimore et al., 2016; Combes et al., 2019; GRAVITY collaboration 2020; 2021), that is, a scale close to the BH sphere of influence. This can be parameterized as the BH Bondi radius $R_B\equiv\frac{GM_{BH}}{c_\infty^2}=\frac{c^2}{2c_\infty^2}r_S$, where $c_\infty$ is the sound speed well beyond $R_B$, and the latter equality gives $R_B$ in units of the Schwarzschild radius $r_S$. $c_\infty=(\frac{\gamma kT}{\mu m_p})^{1/2} \approx 400$km/s when the Compton temperature of the gas irradiated by the AGN is $T_C=2\times 10^7$K, while $\mu$ is the the mean atomic mass per proton, which we fixed to 1.2 (see, e.g., Gofford et al. 2015). Thus, $R_B=2.8\times 10^5 r_S$, that is, 2.7pc for $M_{BH}=10^8 M_\odot$. Therefore, it is plausible to assume that the relativistic wind propagates nearly freely until it impacts the matter around the BH sphere of influence, where the wind thrust may efficiently contrast gas accretion within the Bondi radius. If this is the case,

\begin{equation}
    \dot M_{\rm w} v_w = 4\pi r^2 P_b - 1/f(\bar \omega) \frac {GM_{BH} M_{gas}} {r^2},
\end{equation}
\noindent
where $P_b$ is the thermal pressure in the shocked-wind bubble, and $f(\bar \omega)$ is the fraction of the solid angle covered by the accreting matter that is intercepted by the wind. The thermal pressure in the wind bubble is related to the thermal energy injected by the wind, $E_b = {3/2 P_b V_b}$, where $V_b=4/3\pi r^3$ is the bubble volume, and it can be expressed as $E_b\approx1/2\,L_w \Delta t$ (Weaver et al. 1977). Following the standard Eddington argument, the gas mass that can be just supported by the wind thrust is the BH mass-accretion rate times a characteristic activity time $\Delta t$,  that is, $M_{gas}=\dot M_{accr} \times \Delta t$. Therefore, eq. 12 can be rewritten as follows:
\begin{equation}
\begin{split}
    \dot M_{\rm w} v_w = \frac{\dot M_{\rm w} v_w^2}{2} \frac {\Delta t} {r} - 
     1/f(\bar \omega) \frac{GM_{BH} M_{gas}} {r^2} = \\
    \frac{\dot M_{\rm w} v_w^2}{2} \frac {\Delta t} {r} - 1/f(\bar \omega)\frac{GM_{BH}\dot M_{accr} \Delta t}{r^2} = \\ 
     \frac{\dot M_{\rm w} \Delta t}{r} \times(\frac{v_w^2}{2} - 1/f(\bar \omega)\frac{GM_{BH}}{\bar \omega r} ) .
    \end{split}
\end{equation}
\noindent
$\dot M_{\rm w}$ can be removed from the equation, which can be solved for $\Delta t$,
\begin{equation}
\begin{split}
\Delta t = \frac{f(\bar \omega)v_w r}{f(\bar \omega)\frac{v_w^2}{2}  - \frac {GM_{BH}}{\bar \omega r}} .
\end{split}
\end{equation}
\noindent
When the radius at which the wind impacts with the accreting matter is the Bondi radius, and when we recall that in MHDW systems, $v_w=\frac{\sqrt{2}v_0}{\sqrt{\bar \omega}}$, the equation reads

\begin{equation}
\begin{split}
  \Delta t = \frac{R_B}{\frac{\sqrt{2}v_0}{2 \sqrt{\bar \omega}} - \frac {c_\infty^2}{\sqrt{2}v_0 f(\bar \omega) \sqrt{\bar \omega}}} \\
  = \frac{R_B \sqrt{\bar \omega}}{0.7 v_0 - \frac{c_\infty^2}{\sqrt{2}v_0 f(\bar \omega)}}.
\end{split}
\end{equation}
\noindent

The second term of the denominator is negligible unless $f(\bar \omega)<<1$, thus $\Delta t \approx 140\;{\rm yr} \times \sqrt{\bar \omega} (\frac{v_0/c}{0.1})^{-1}$ if $M_{BH}=10^8 M_\odot$ and $R_B=8.3\times10^{18}$cm. This is similar to the timescale derived by King \& Pounds (2015), which again suggests that many wind/accretion episodes can easily occur during a typical AGN lifetime of several millions years. 

$\Delta t$ depends on $\sqrt{\bar \omega}$. We recall that $\lambda=\lambda_w/\bar \omega$ and that $\bar \omega <1$. This therefore suggests that high $\lambda$ states are shorter than low $\lambda$ states. When a source spends a shorter time in high $\lambda$ state than in lower $\lambda$ states, the probability of finding this source in the former state is lower than the probability of finding it in a state of low $\lambda$. Consequently, the "observed" $\lambda$ distribution in a blind survey is skewed toward low $\lambda$ values (the most likely states). Thus, the observed $\lambda$ distribution should be compared with the expected distribution, weighted by the different times the sources remain in any given $\lambda$ states, that is, $\lambda \times \Delta t \propto \frac{\lambda_w}{\sqrt{\bar \omega}}$. The black curve in Figure \ref{lamdistr} represents the distribution of $\lambda \times \Delta t = \frac{\lambda_w}{\sqrt{\bar \omega}}$. The slope of the $\lambda \times \Delta t$  distribution above 0.05 after correcting current UFO measurements for selection biases is $-0.9\pm0.5$, which is consistent with both the Bongiorno et al. (2012) determination, $-0.96$ in the redshift range 0.3-0.8 and Aird et al. (2012;2018) determination, $-0.65$. Ananna et al. (2022) estimated the $\lambda_{Edd}$ distribution in a sample of local AGN and reported a broken power law with a break $\lambda_{Edd}$ similar to our $\lambda_w$ threshold, but with a a slope above the break of $\geq -2$, which is steeper than that of the $\lambda \times \Delta t$ distribution and that of previous determinations. Kauffmann \& Heckman (2009) reported log-normal distribution functions due to the presence of low-luminosity AGNs in their samples.
 
We finally computed the total BH growth obtained by a large number (10,000-100,000) of accretion events with $\lambda$ following the distribution function in Figure \ref{lamdistr} (the black histogram shows the constant time spent in each accretion episode). This is equivalent to the BH growth obtained at a constant $\lambda\sim0.1$ during the entire AGN activity time. This number is very sensitive to details of the $\lambda_w$ and $\bar \Omega$ distributions: their minimum value, the inflection points, and the distribution slopes. Therefore, the above figure should not be considered a firm limit for real accreting systems. We limit ourselves to noting that if the three main assumptions of the above analysis are correct (1. a link between accretion and wind emission, i.e., $\dot M_{accr}=1/\bar \omega \dot M_{\rm w}$; 2. an Eddington-like limit on the mass-accretion rate because the wind thrust contrasts the accretion flow at the Bondi radius; and 3. a random distribution of accretion/wind events), then systems can hardly accrete close to their Eddington rate for the full AGN time cycle.  High accretion rate events, even when they exceed several Eddington units, are likely, but their number will be small in comparison to low accretion rate episodes and their duration will be shorter. 

\section{Is the disk-wind system a self-organized critical system?}

Self-organized criticality (SOC) was first introduced by Bak et al. (1987, 1994). It is a characteristic of some complex dynamical system. These systems reach the critical state not by tuning an external parameter, but because of the self-organization of internal interactions, among which feedback is the most important. SOC systems develop power-law scaling laws and are therefore scale invariant. Because of these intriguing similarities with disk-wind systems (natural development of feedback and power-law scaling laws), it is reasonable to wonder whether disk-wind systems are SOC systems. The first pioneering work on SOC in BH astrophysics (Mineshige et al. 1994a,1994b, 1999) followed the introduction of SOC systems in the literature by just a few years. These works were focused on interpreting the chaotic X-ray variability of stellar BH and AGN in the framework of SOC models. Mineshige and collaborators were successful in generating the typical 1/f-like fluctuations that are often observed in these sources using a simple cellular automaton based on simple SOC rules. Their conclusions were later questioned by Uttley et al. (2005) and Done et al. (2007) on the basis that SOC models naturally produce power-law distributions of the temporal fluctuations, while log-normal distributions are typically observed in stellar BH systems. In AGN, the situation is not clear cut because in the best-studied case, that of the narrow-line Seyfert 1 IRAS 13224-3809 (Alston et al. 2019), there are strong signs of nonstationarity and strong departures from a log-normal distribution of fluxes. Even the presence of power-law distributions is not sufficient to assess that the system can be described by the theory of critical phenomena, however, because many mechanisms can be at work to generate a power-law distribution. In addition to the power spectrum of the time series of the events, SOC systems must show the typical relation between some scaling exponents (Kuntz \& Sethna 2000; Sethna et al., 2001). At least three other fundamental scaling laws must be analyzed: the histograms of the size $S$ and duration $T$ of the burst $f(S)\sim S^{-\tau}$, $f(T)\sim T^{-\tau_t}$, and the scaling of the average size of avalanches with a given duration $<S>(T)\sim T^{\alpha_1}$. The parameters $\tau, \tau_t$, and $\alpha_1$ are critical exponents of the system, regardless of the microscopic details of the system. Scaling theory (Kuntz \& Sethna 2000; Sethna et al., 2001) predicts exponent relations, and in particular, 
\begin{equation}
\alpha =   \frac{(\tau_t - 1)}{(\tau - 1)}, 
\end{equation}
\noindent
when the system is at criticality, $\alpha=\alpha_1$. All critical exponents can be measured both in data streams and in models, and the above relation allows us to assess whether the system is consistent with an SOC state. 

We first reproduced the Mineshige et al. (1994) cellular automaton according to the following steps:
\begin{enumerate}
\item We divided the disk plane into cells along the radial and azimuthal coordinates. We chose $N_\phi=128$ azimuthal segments and $N_r=64$ radial segments (see the discussion below for a justification of this choice).

\item We randomly chose one cell in the outermost ring and placed an accreting gas particle of mass $m$ at each time step. $M_{i,j}$ is the mass of a cell at radius $r_i$ and azimuthal coordinate $\phi_j$.

\item We verified whether the mass accumulated at any location exceeded a given threshold, that is, whether $M_{i,j}>M_{crit}$. The results do not depend on the actual value of this threshold. We started the simulation by assigning a mass to all disk cells. We verified that the results did not change for different starting conditions, for example, adopting the condition of an initial empty disk, adopting a random distribution of cell masses, or starting by assigning a mass to all disk cells of $M_{crit}$. Every time $M_{i,j}>M_{crit}$ , three mass particles fell from the cell into adjacent cells at $r_{i+1}$, that is, an accretion event occurred in the disk,
    
\begin{equation}
\begin{split}
    M_{i,j}=M_{i,j}-3m\\
    M_{i+1,j}=M_{i+1,j}+m\\
    M_{i+1,j+1}=M_{i+1,j+1}+m\\
    M_{i+1,j-1}=M_{i+1,j-1}+m\\.
\end{split}
\end{equation}

Given the circular boundary condition in 1), if $j=128$ $j+1=1$, and if $j=1$ $j-1=128$ in the above expressions. This accretion of mass may cause that the mass in some of the inner $r_{i+1}$ cells might also exceed the critical threshold, and an avalanche thus propagates inward at the following time steps. We call an avalanche the propagation of an accretion event through the disk.  

\item We repeated the procedure $n$ times to obtain a statistical population from which temporal and spatial distributions can be drawn.
\end{enumerate}

Minishege et al. (1999) introduced an additional role to account for inward diffusion, in order to better match observed variability light curves. We did not consider diffusion here because we are not interested in reproducing the fast disk variability, but rather the long-term disk evolution. 

In order to describe the statistical properties of the system, two main quantities were calculated for each avalanche: (a) the lifetime of the avalanche $T$, that is, the number of time steps of single avalanche propagation, and (b) the size of the avalanche $S$, that is, the total number of accretion episodes during the avalanche evolution. At each time step, several avalanches can be active on the disk. We finally followed Mineshige et al. (1994) in calculating the total energy that is liberated at each time step (adding the contributions from each active avalanche).

For our setup, the maximum avalanche size was limited to 4096 ($\sum_i^{64} (i*2)-1$). This largest avalanche covered all 128 cells in the innermost ring. We therefore set the number of azimuthal elements ($N_r\times2$). The original setup of Mineshige et al. (1994, 1999) used $N_\phi=N_r=64$, implying that avalanches propagating inside $r=32$ spread multiple masses on the same cell. This additional accumulation of mass on some cells with respect to the simple rule n. 3 above was counterbalanced by assuming an $M_{crit}$ proportional to $r$. Using the setup in step (1) above, we dropped this additional complexity and used a constant $M_{crit}$. We were able to reproduce the results of Mineshige et al. (1994) using both our simple setup and their original setup with $M_{crit}$ proportional to $r$. We verified that the results did not depend on the actual value of $M_{crit}$ provided that $M_{crit}\geq3m$.

We ran this simple cellular automaton to build the distributions $f(T)$ and $f(S)$. To reduce the noise due to the limited number of avalanches with large sizes, which hinders an assessment of the real shape of the distribution $f(S)$ for high S values, we ran the cellular automaton to produce $>10$ avalanches of size $S>1000$. We produced a total of 210.000 avalanches and discarded the first 10.000 to remove possible initial transients.

\begin{figure*}
\vspace{-0.5cm}
\begin{tabular}{ccc}
\includegraphics[width=5.7cm]{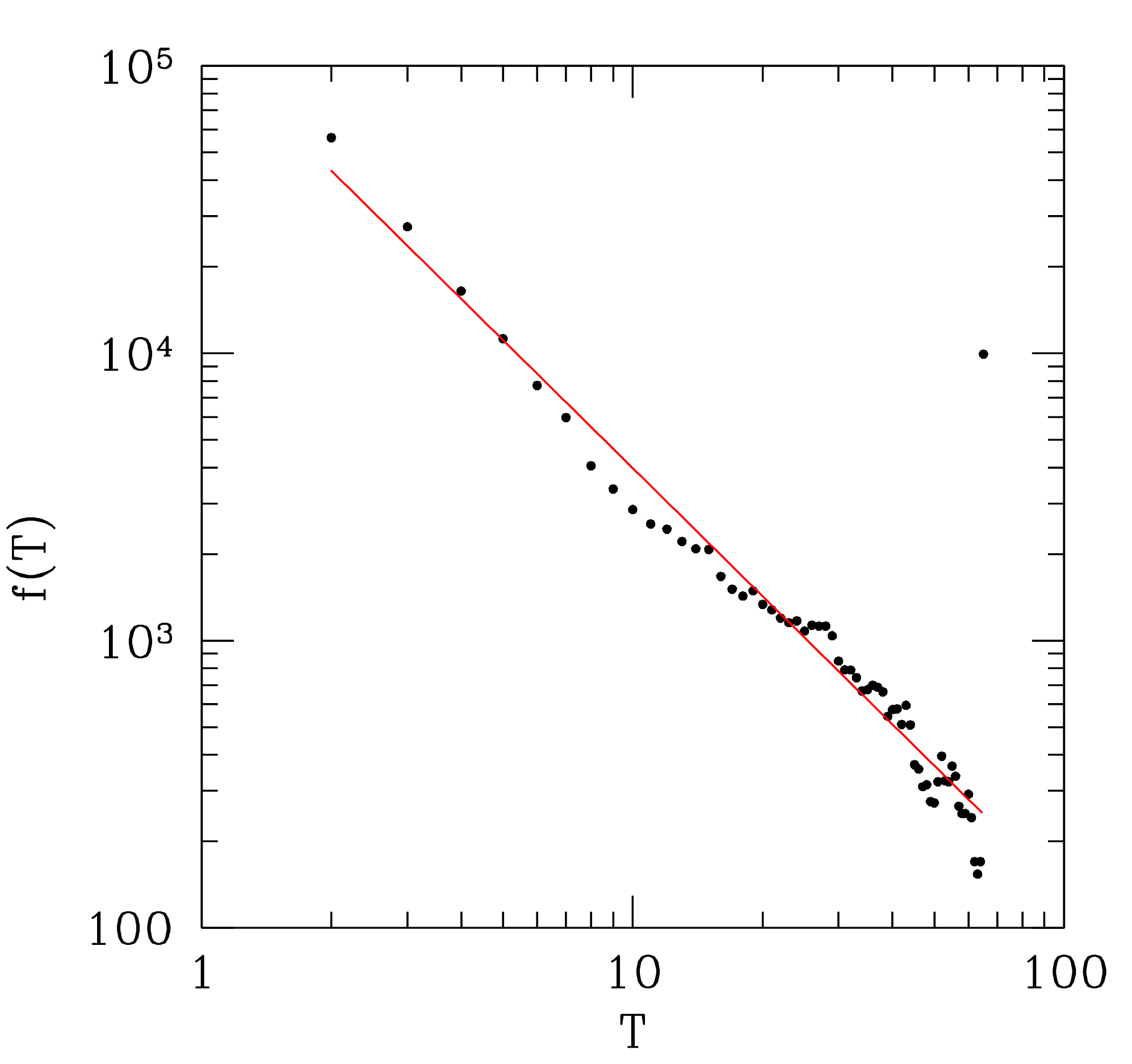}
\includegraphics[width=5.7cm]{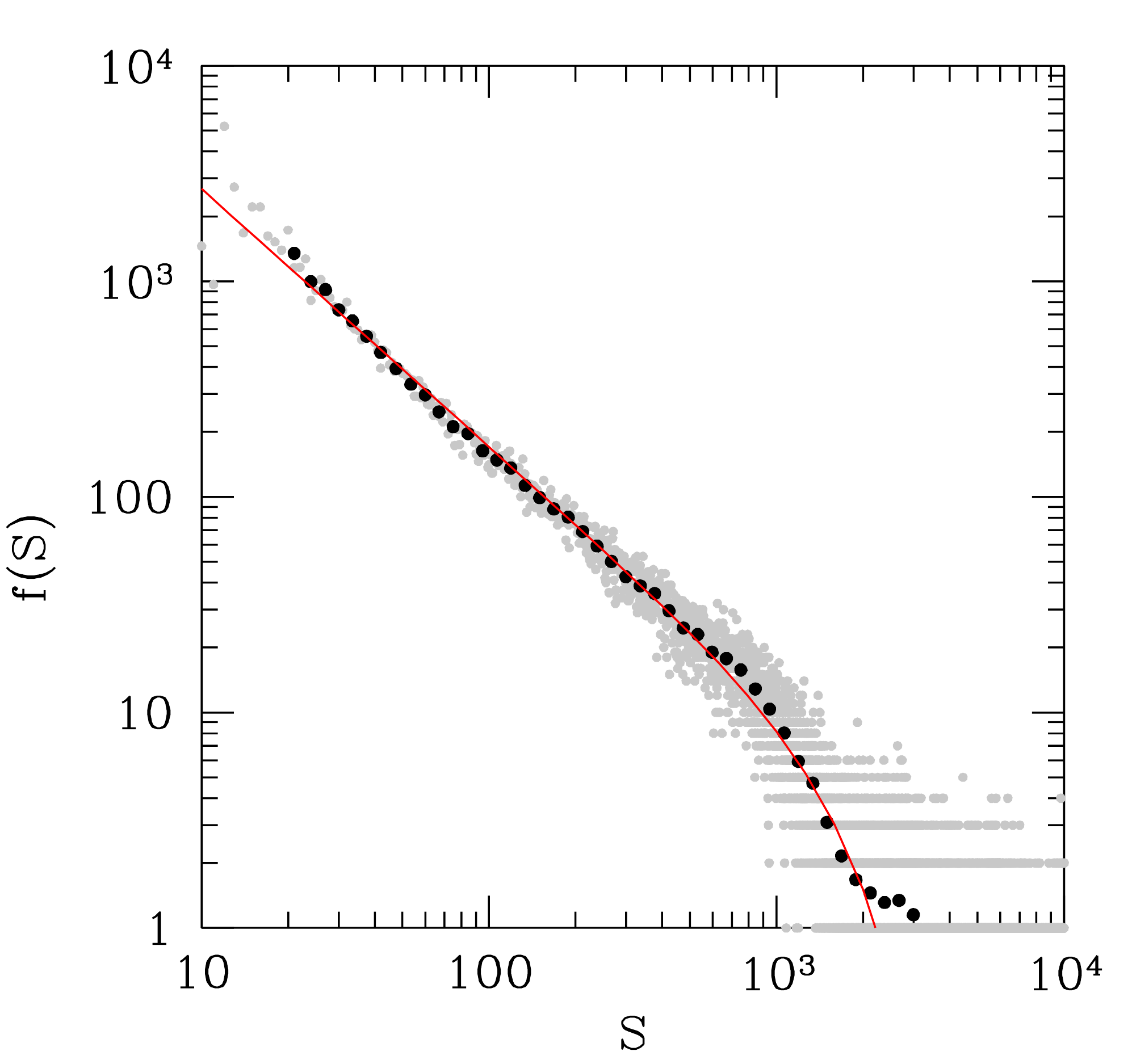}
\includegraphics[width=5.7cm]{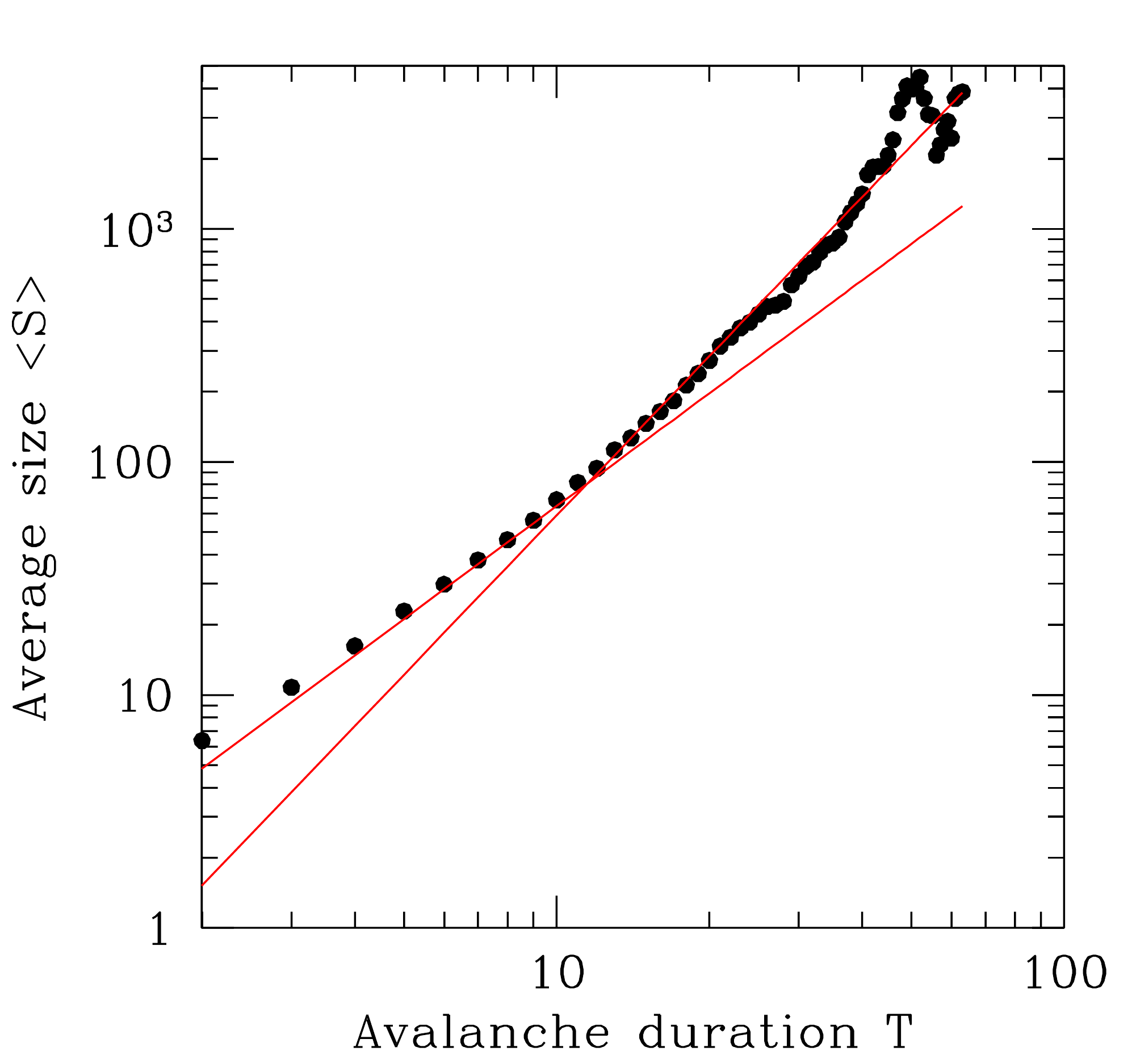}
\end{tabular}
\begin{tabular}{ccc}
\includegraphics[width=5.7cm]{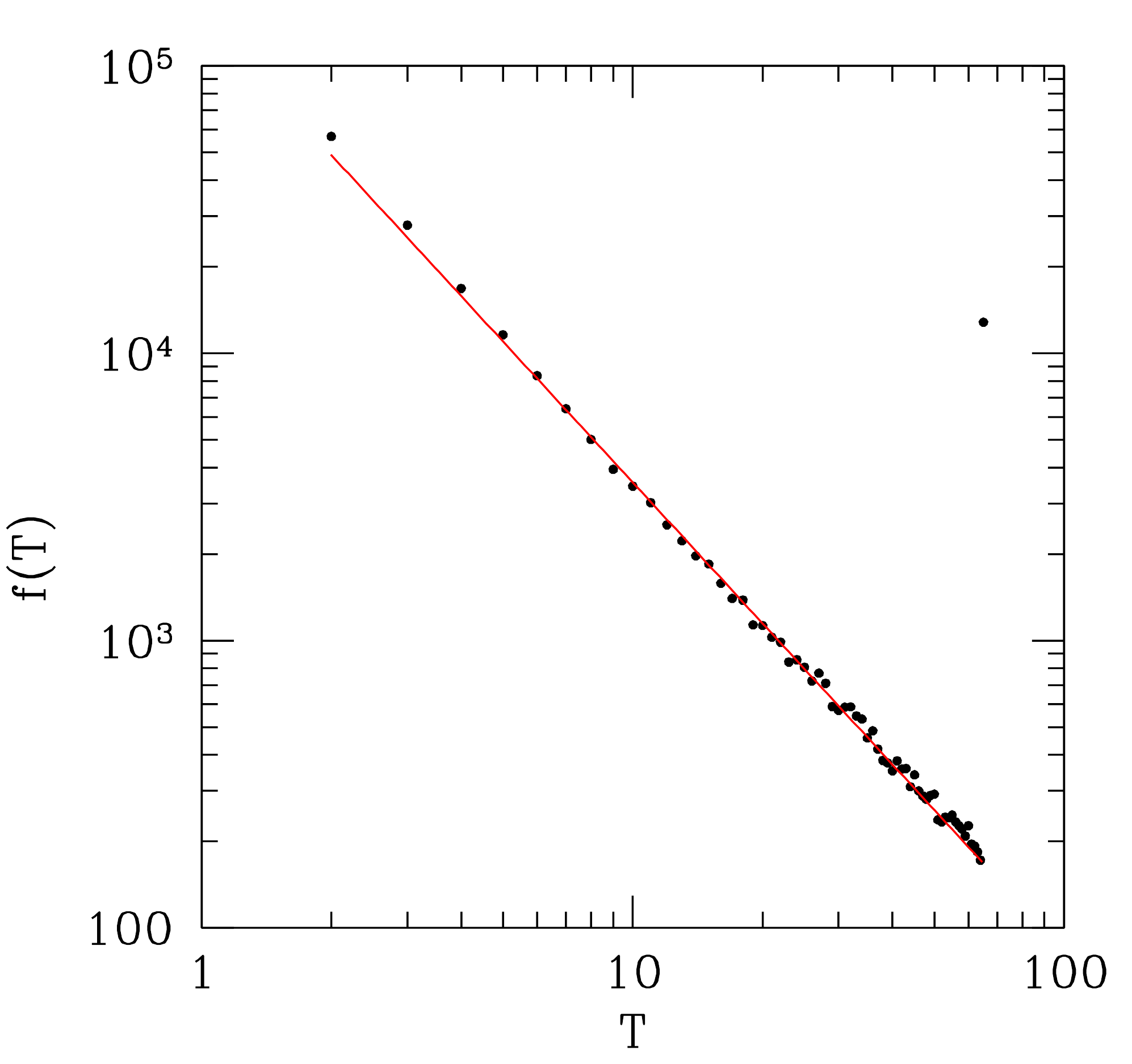}
\includegraphics[width=5.7cm]{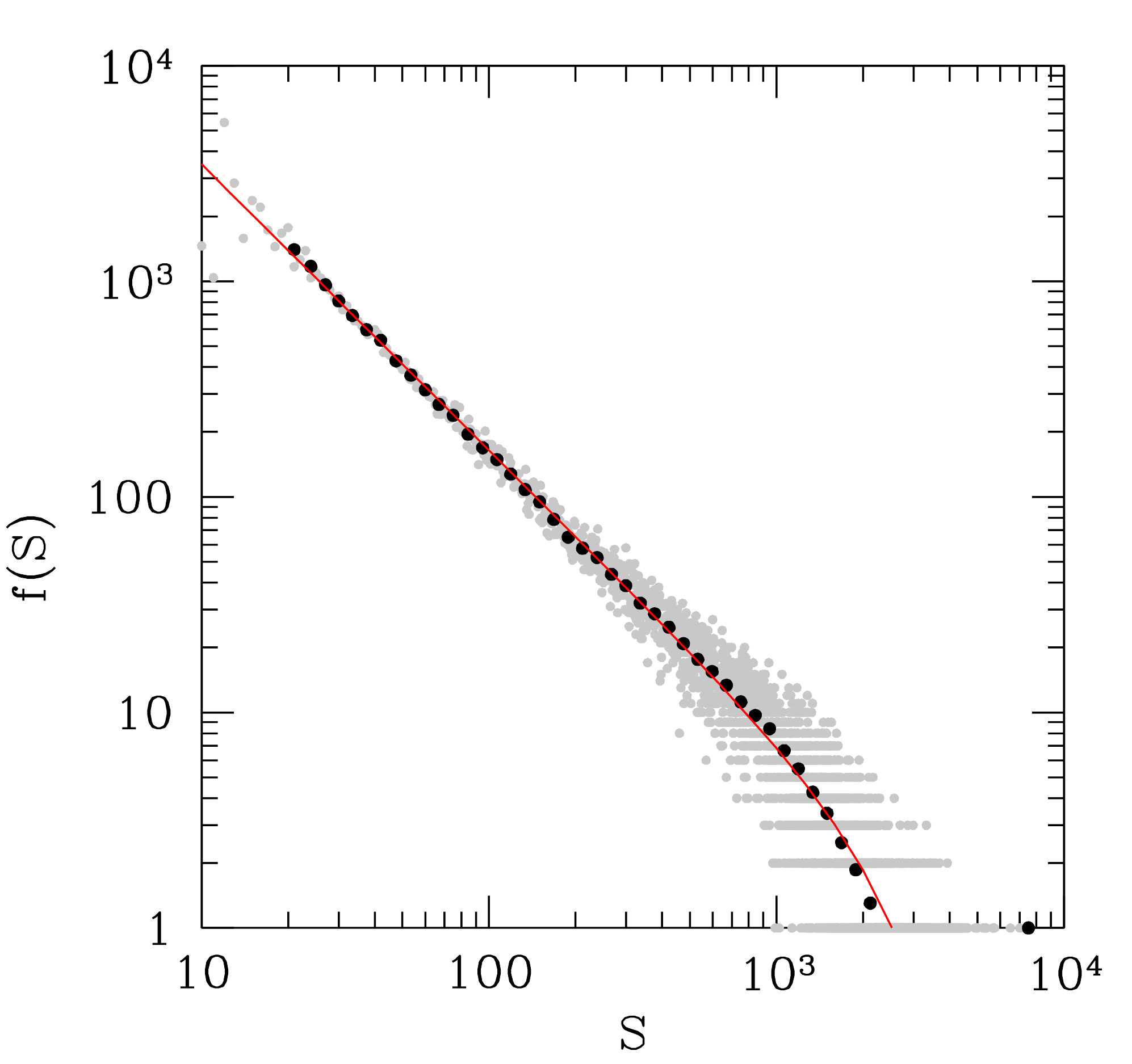}
\includegraphics[width=5.7cm]{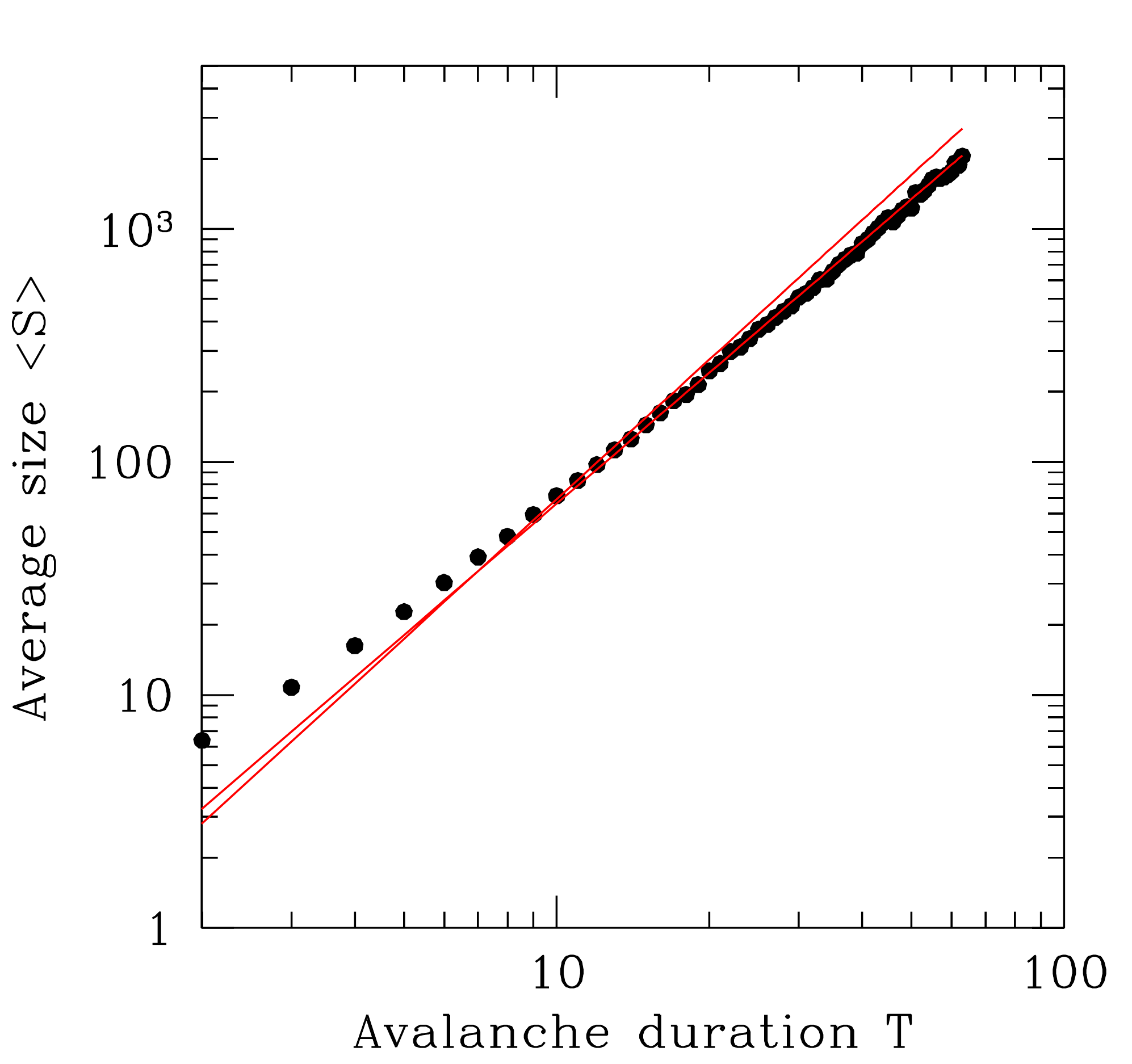}
\end{tabular}
\begin{tabular}{ccc}
\includegraphics[width=5.7cm]{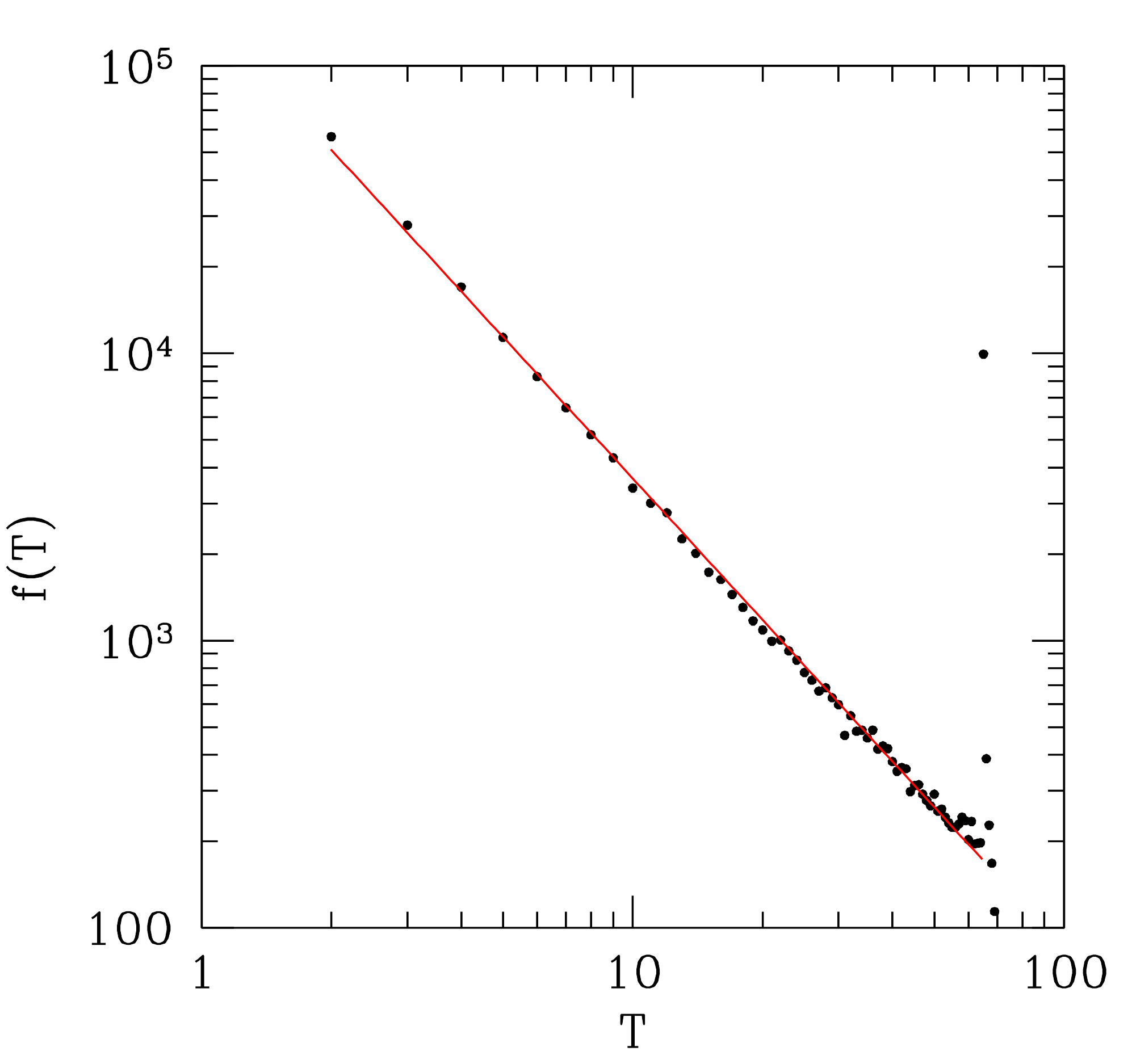}
\includegraphics[width=5.7cm]{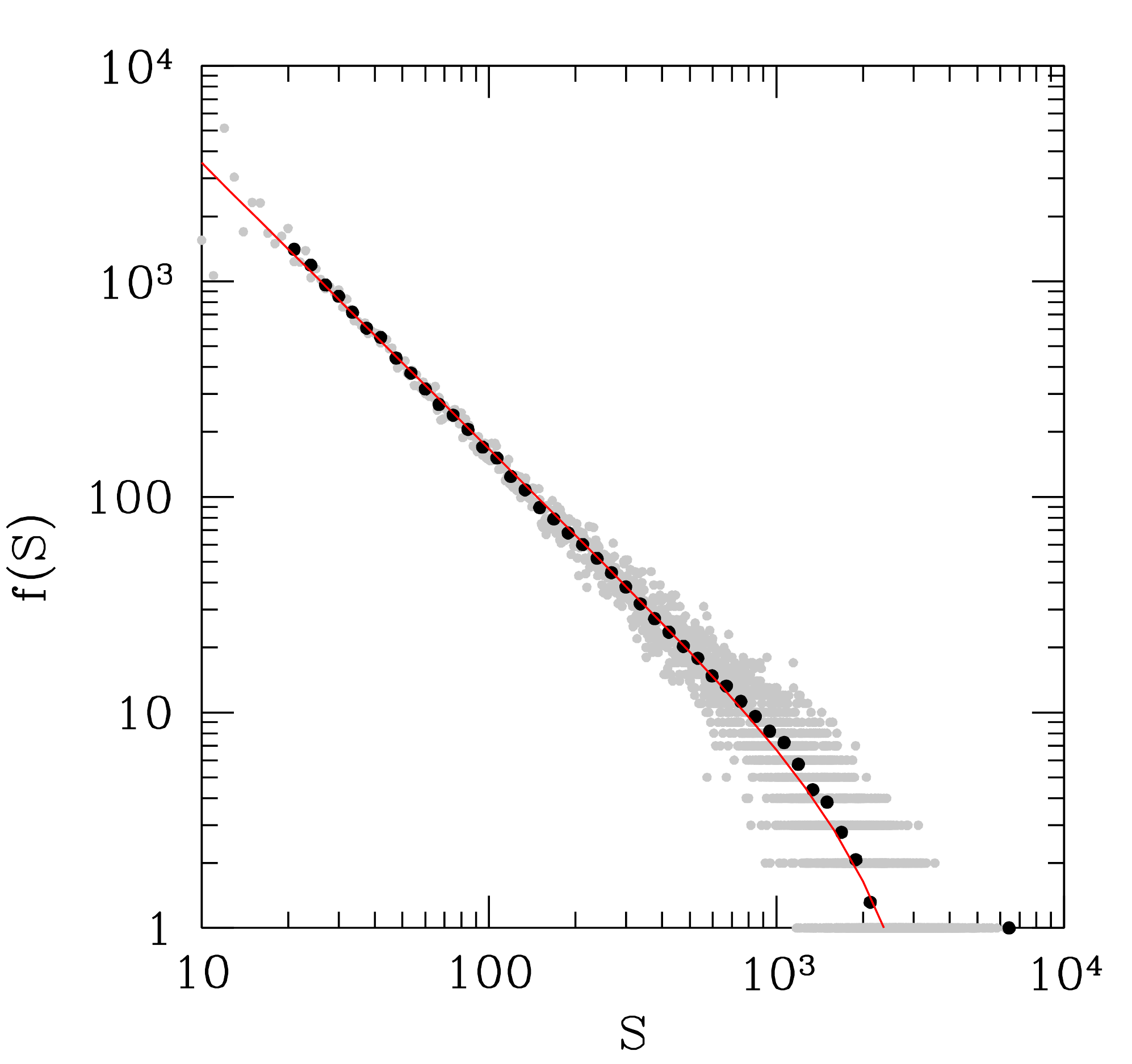}
\includegraphics[width=5.7cm]{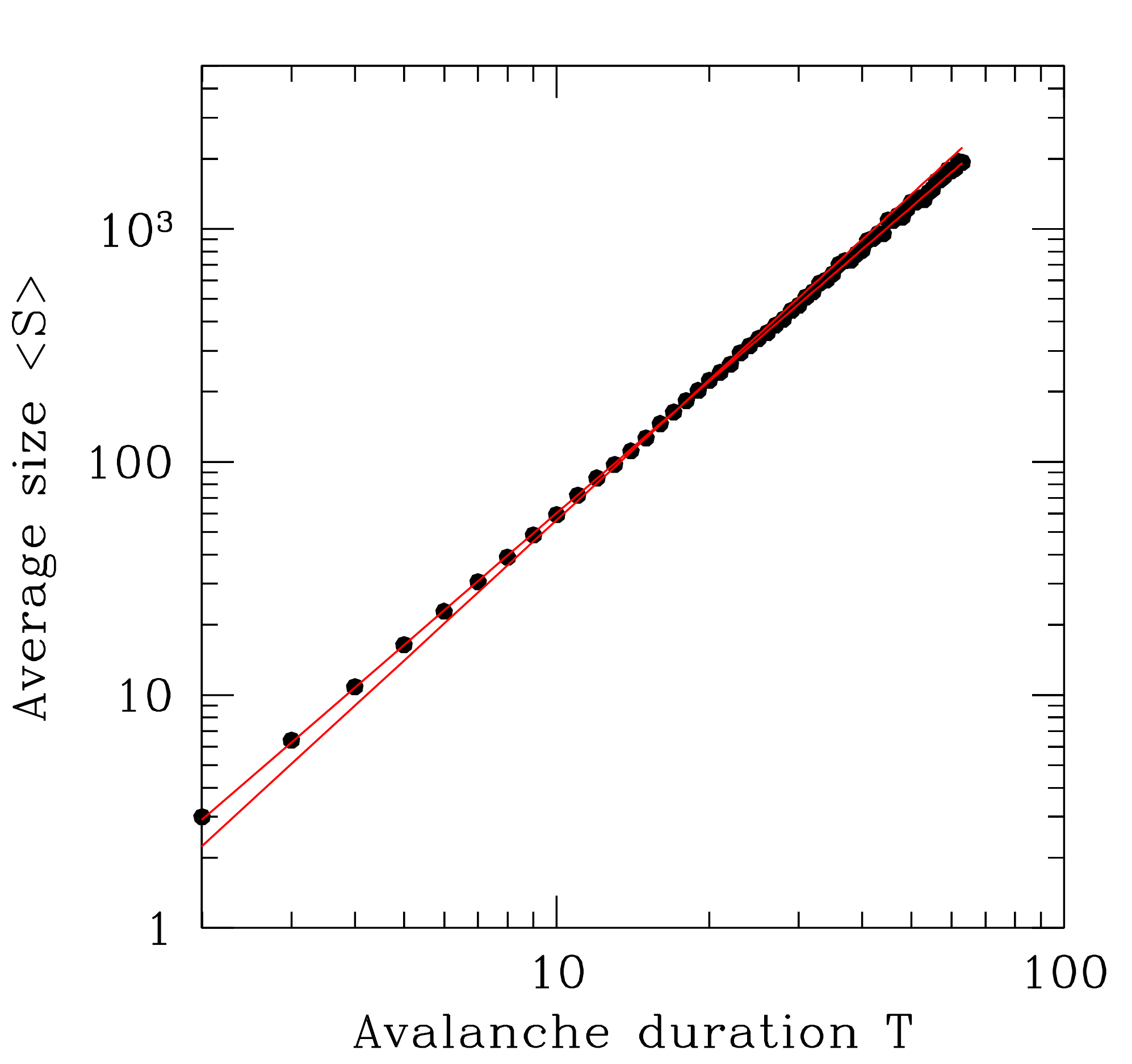}
\end{tabular}
\begin{tabular}{ccc}
\includegraphics[width=5.7cm]{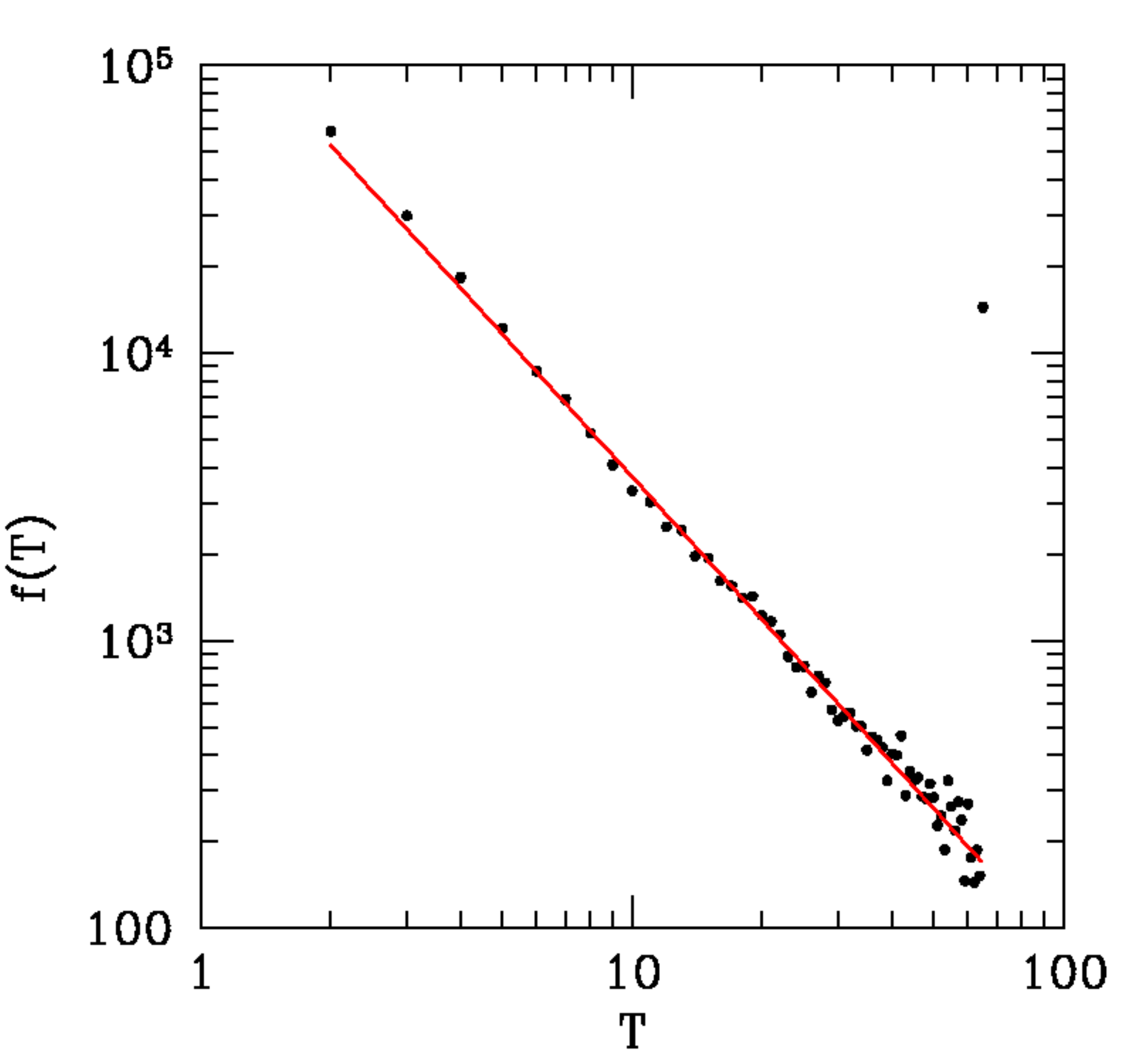}
\includegraphics[width=5.7cm]{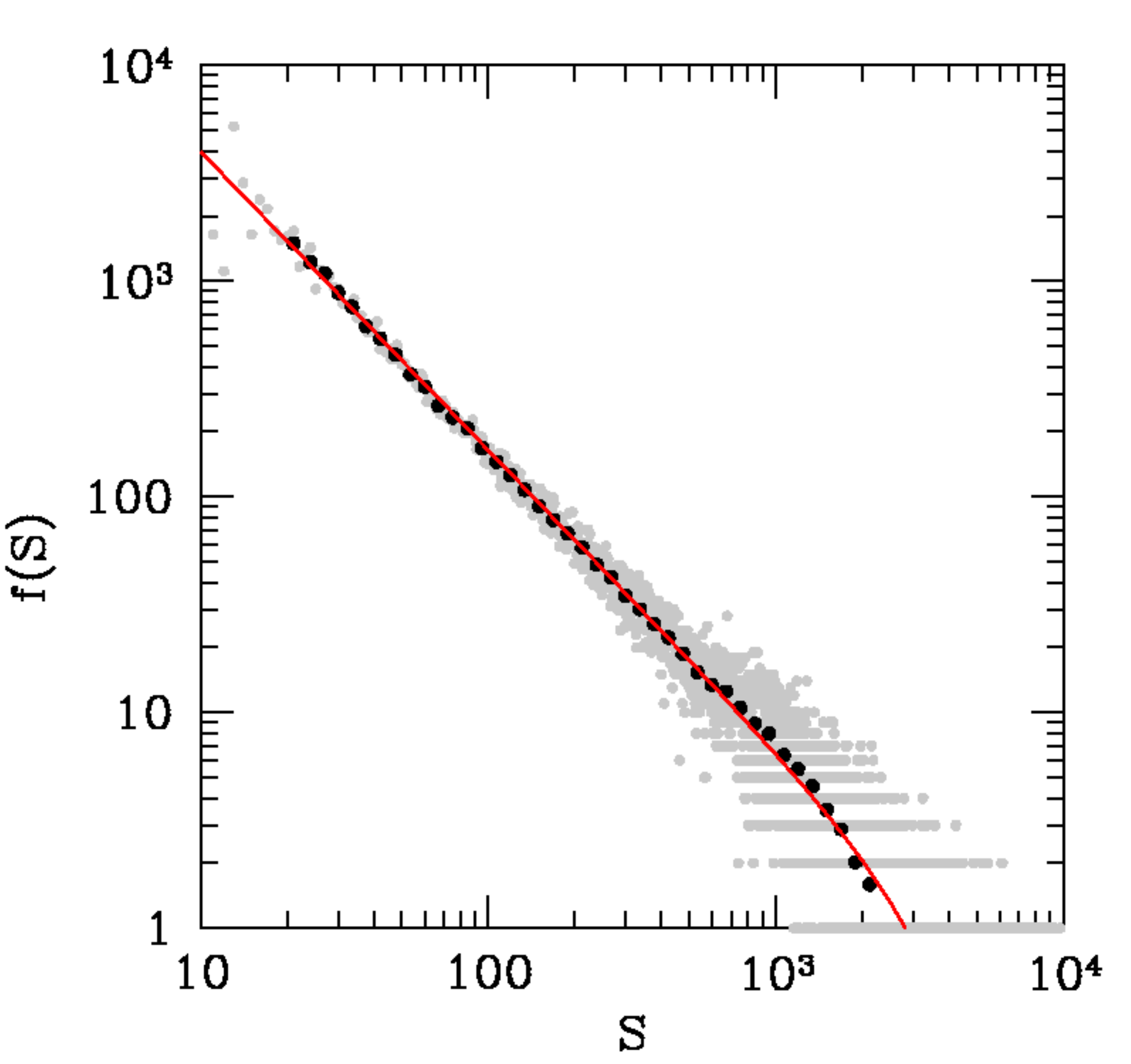}
\includegraphics[width=5.7cm]{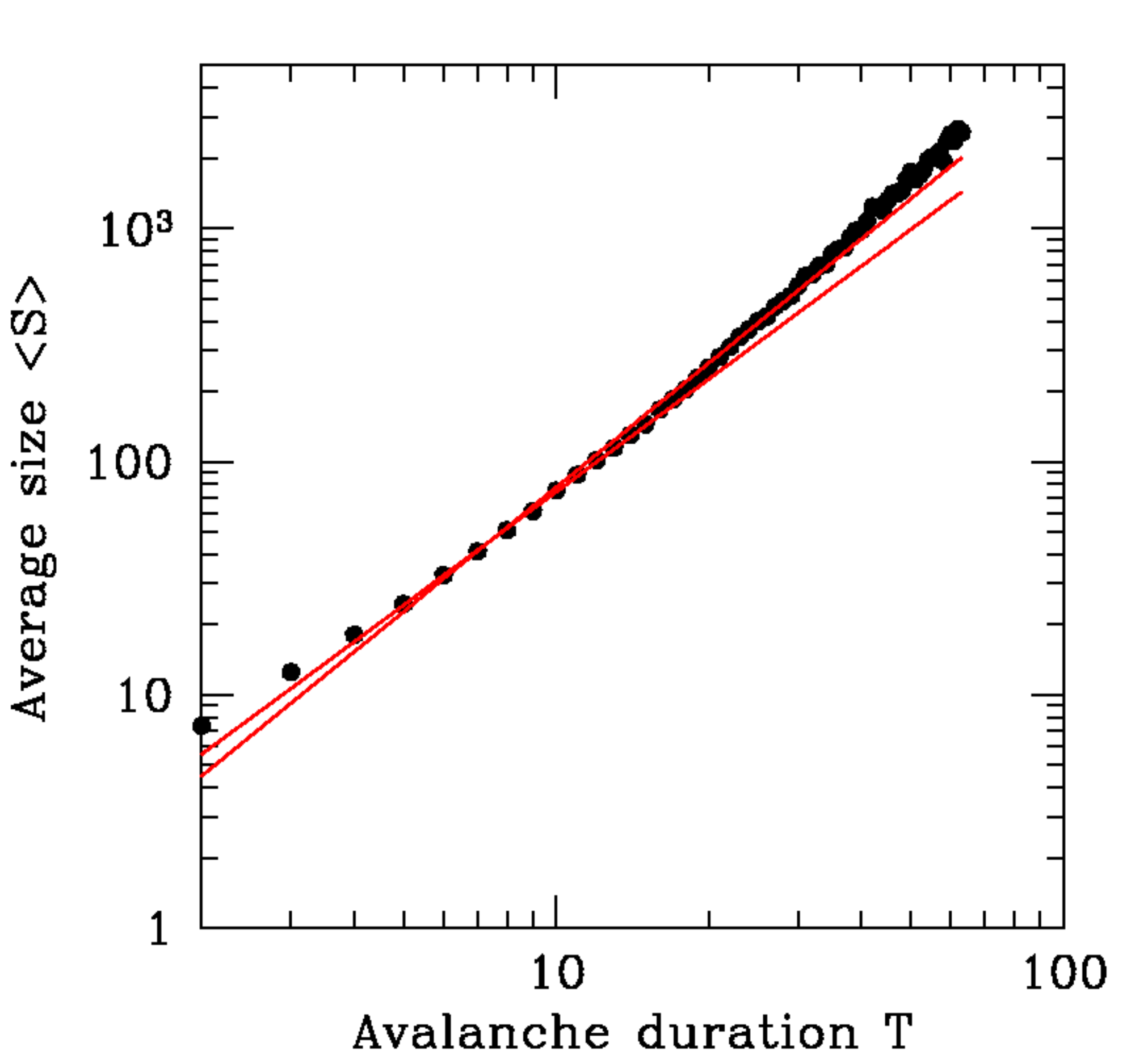}
\end{tabular}
\vspace{-0.5cm}
\caption{[Left panels]: Distribution of the avalanche duration $f(T)$. [Central panels]: Distribution of the avalanche size $f(S)$. The gray points are the original distribution function, and the black points mark the distribution logarithmically binned.  [Right panels]: Average avalanche size $<S>$ vs. avalanche duration $T$. [Upper panels]: Setup 1, similar to the original simulation of Mineshighe et al. (1994). [Central upper panels]:  Setup 2, including a link between mass accretion and wind mass outflow as in MCDW models. [Central lower panels]: Setup 3, including a link between mass accretion and wind mass outflow as in MCDW models and an Eddington-like limit on the mass-accretion rate due to the wind thrust that contrasts the accretion flow at the first radius.  [Bottom panels]: Setup 4. This is the same as setup 1, but includes a threshold on the disk luminosity above which the accretion is halted (similar to an Eddington limit). The red lines in the left and central panels are best-fit models. The red lines in the right panels bridge the interval of allowed $\alpha$.}
\label{ST1}
\end{figure*}

The upper panels in Figure \ref{ST1} show the distributions of $f(T)$, $f(S)$, and the average size for a given duration $<S>(T)$. Table \ref{tab:ST} lists the slope $\tau_t$ of the $f(T)$ distribution, the slope $\tau$ of the $f(S)$ distribution, the slope $\alpha_1$ of the $<S>$ versus $T$, and $\alpha=\frac{(\tau_t - 1)}{(\tau - 1)}$. $\tau_t$ and $\tau$ were obtained by fitting the $f(T)$ and $f(S)$ distributions with a power law plus an exponential-cutoff model. This cutoff is due to the limited size of the system and moved to higher S when systems with $N_r>64$ and $N_\phi>128$ were considered. The point at $T=65$ in the $f(T)$ distribution represents the avalanches exiting the system from the inner radius. The best-fit $\tau_t$ and $\tau$ are close to the values of -3/2 and -4/3 that are expected for variants in the original Bak et al. (1987) model, introducing a preferential direction (Ben Hur et al. 1996). The errors on $\tau_t$ and $\tau$ are dominated by systematic uncertainties, that is, the wiggle in the $f(T)$ distributions and the excess at high $S$ in the $f(S)$ distribution for setup 1. These wiggles suggest that the system is supercritical and therefore out of SOC. This is further supported by the mismatch between $\alpha=\frac{(\tau_t - 1)}{(\tau - 1)}$ and $\alpha_1$ (see Table \ref{tab:ST}). 

\begin{table*}[]
    \centering
    \begin{tabular}{lccccc}
    \hline 
 & $\tau_t$ & $\tau$ & $\alpha_1$ & $\alpha$ & $\alpha_{coll}$\\
\hline
(1)  & -1.46$\pm$0.04    & -1.24$\pm$0.03     & 1.45 - 2.25 & 1.92$\pm$0.08 & $\sim2$ \\
(2)  & -1.630$\pm$0.013 & -1.327$\pm$0.010   & 1.49 - 1.82 & 1.93$\pm$0.06 & 1.92$\pm$0.02\\
(3)  & -1.636$\pm$0.021 & -1.327$\pm$0.010   & 1.89$\pm$0.02 & 1.94$\pm$0.06 & 1.92$\pm$0.02\\
(4)  & -1.650$\pm$0.015 & -1.385$\pm$0.010   & 1.27 - 1.98  & 1.69$\pm$0.06 &  2.15$\pm$0.03\\ 
    \end{tabular}
    \caption{Avalanche exponents for the different setups. Setup 1: Baseline cellular automaton, following Mineshige et al. 1994. Setup 2: Cellular automaton with a link between gas accretion and wind emission, as in MCDW models. Setup 3: As in setup 2, but with an Eddington-like limit on the mass-accretion rate at the outer radius.  Setup 4: As in setup 1, but with a threshold on disk luminosity above which the accretion is halted. $\alpha_1$ is the best linear regression to the full $<S>(T)$  distribution for setup 3, while for setups 1, 2, and 4 the two values refer to the best linear regression for $T<10$ and $T>10$.} 
    \label{tab:ST}
\end{table*}

We recall that setup 1 is conceptually similar the original sand-pile model of Bak et al. (1987), with the introduction of a preferential direction. In the experiment of Bak et al. (1987), SOC is achieved through self-regulation between the mass entering the system and the mass exiting the system boundaries. In our implementation, the possibilities for the mass to exit the system are severely reduced with respect to the original experiment because the only egress is via the central compact object. This produces an excess in the number of avalanches close to the maximum size. 

Setup 1 does not include feedback processes, and we considered it as a benchmark against which to compare results arising from more complex scenarios. To gain further insights, we added three additional ingredients to the baseline cellular automaton: a link between gas accretion and wind emission, as in MCDW models (setup 2), an Eddington-like limit on the mass-accretion rate due to the wind thrust that contrasts the accretion flow at the outer radius (setup 3), and a modification of setup 1 to account for the Eddington limit to the gas accretion due to radiation pressure (setup 4) as in RDW.

In setup (2), at each location of the disk where $M_{i,j}>M_{crit}$, we randomly extracted a poloidal Alfv\'en velocity in units of the Keplerian velocity at the magnetic wind footpoint $v_{K0}$ from 0.01 to 100. This is a proxy for the poloidal magnetic field. We then used the relations in Figure 3 of Cui \& Yuan (2019) to estimate $r_A/r_0$ and thus $\bar \omega=(r_A/r_0)^{-2}$. The fiducial model of Cui \& Yuan (2019) is a hot-accretion flow, but the authors discussed deviations with respect to this baseline. They discussed that winds emerging from thin disks may have different properties. In particular, thin disks are cooler and thus have lower sound speeds $C_{s0}$ than thick disks. For this reason, the authors calculated models not only for $C_{s0}=0.5 v_{K0}$, as appropriate for hot disks, but also for $C_{s0}=0.1 v_{K0}$, as appropriate for cold disks. We used both models from Figure 3 of Cui \& Yuan (2019) in our cellular automaton, and the differences were negligible.

In setup (2), the mass transferred to $M_{i+1,j}, M_{i+1,j+1}$ and $M_{i+1,j-1}$ was set to $m_{in}=m/(1+\bar \omega$), while a mass $m_{out}=m\bar \omega/(1+\bar \omega)$ left the system in the form of a wind. The system has a preferential direction, similar to setup 1, but mass can leave the system at each radius, so that not all masses that accrete at the outer radius are forced to reach the inner radius. This is different from both the original Bak et al. (1987) setup and the directional setup of Ben Hur et al. (1996). $\tau_t$ in Table \ref{tab:ST}, is -1.63, steeper than the value expected for directional avalanches. This indicates a different universality class. We find that setup (2) approaches criticality because $\alpha_1\sim \alpha$ (see Table \ref{tab:ST} and Figure \ref{ST1}). However, it must be noted that the scaling of $<S>$ versus $T$ does not follow a pure power-law shape. The best-fit slope for $T<10$ is 1.49, but it is 1.82 for $T>10$. 

\begin{figure*}
\includegraphics[width=16cm]{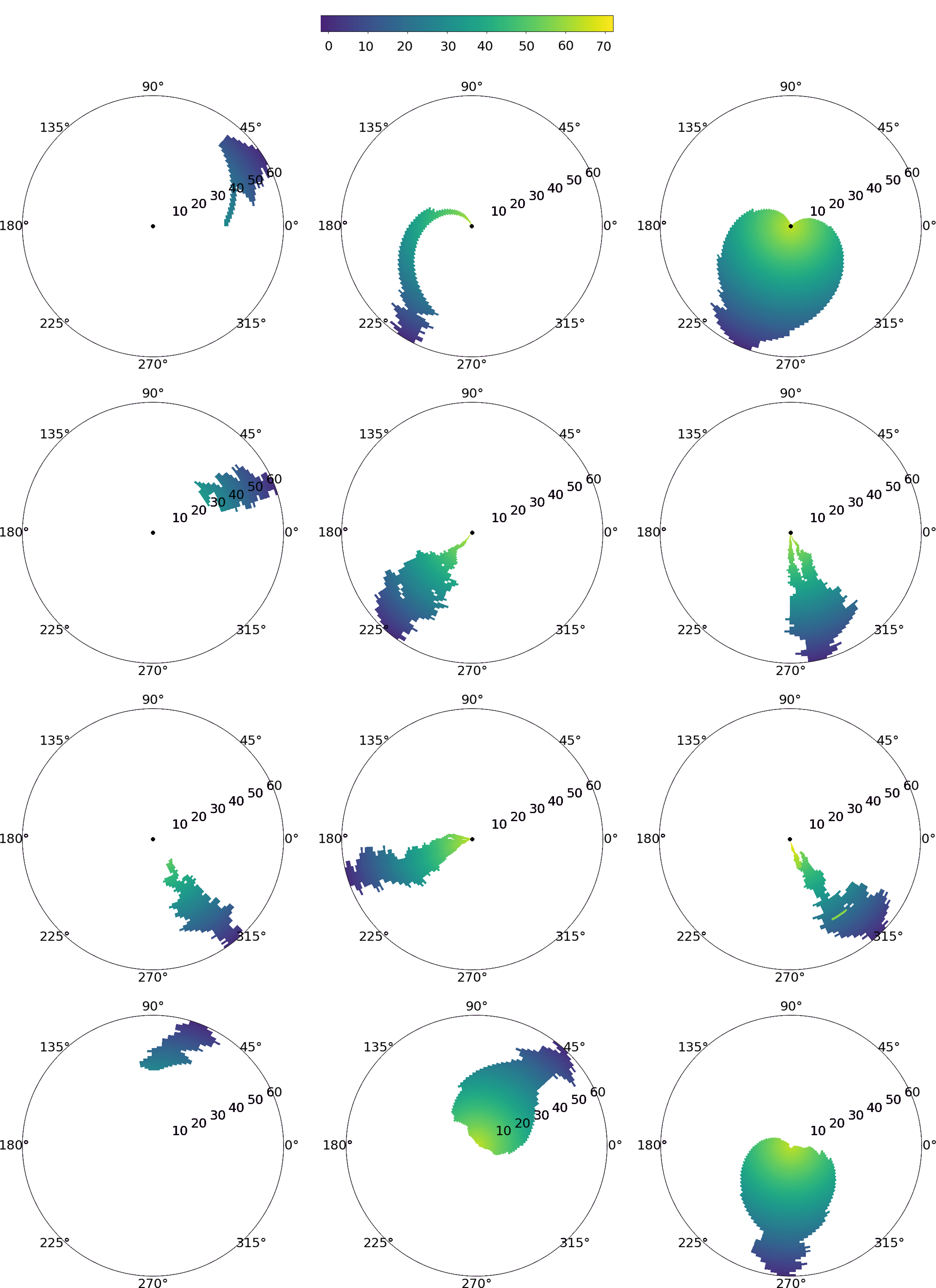}
\caption{Trajectories of three avalanches for each setup (from top to bottom, setups 1, 2, 3, and 4), propagating through the disk. The left panels show avalanches that do not reach the innermost radius, the central panel shows complete avalanches, and the right panel shows the (complete) avalanches with the largest size for that given setup. The central black dot, located at the innermost radius, corresponds to the sink through which the avalanches exit the system. The color-coding indicates the time since the onset of the avalanche. The legend is reported at the top.}
\label{polar_figs}
\end{figure*}

Setup 3 is based on setup 2 and further includes an Eddington-like limit on the mass-accretion rate due to the wind thrust that contrasts the accretion flow at the outer radius. At each time step, the mass that accretes on the outer radius is reduced by a factor $<\bar \omega>$, that is, $m_{in}=m\times (1-<\bar \omega>$) at this location. If $<\bar \omega>$ is small, $m_{in}=m$, while if $<\bar \omega>$ approaches one, there is negligible accretion at the outer radius. This is a rather rough feedback recipe, and it should not be regarded as a realistic disk-wind model. We studied this setup to understand how and in which direction the introduction of this further feedback process modifies the dynamical status of the system. The inclusion of this further feedback brings the system even closer to SOC. In this setup, the scaling of $<S>$ versus $T$ does follow a pure power law.

Finally, we considered a different modification of setup 1 by introducing a threshold in disk luminosity above which the accretion is abruptly truncated. The disk luminosity was calculated following Mineshige et al.~(1994) and assuming that the \textit{k}th gas particle falling from the ring {\it i} to the ring {\it j} liberates a potential energy proportional to $\frac {1} {r_j} - \frac {1} {r_i}$,
\begin{equation}
    L_{\rm disk}\propto \sum_k (\frac {1} {r_j^k} - \frac {1} {r_i^k}).
\end{equation}
\noindent
If $L_{\rm disk}>L_{\rm threshold}$, the gas particle is pushed outside the system by radiation pressure and the accretion is blocked. This causes the luminosity to decrease, resuming the accretion and starting a new cycle. The inclusion of this feedback regularizes the flow, and the system is much more closed to criticality than for setup 1 because $\alpha_1\sim \alpha$ (see Table \ref{tab:ST} and Figure \ref{ST1}). However, as in setup 2, the scaling of $<S>$ versus $T$ does not follow a pure power law in this case either. The best-fit slope for $T<10$ is 1.27, but it is 1.98 for $T>10$. 

\begin{figure*}
\begin{tabular}cc
\includegraphics[width=8.0cm, height=8cm]{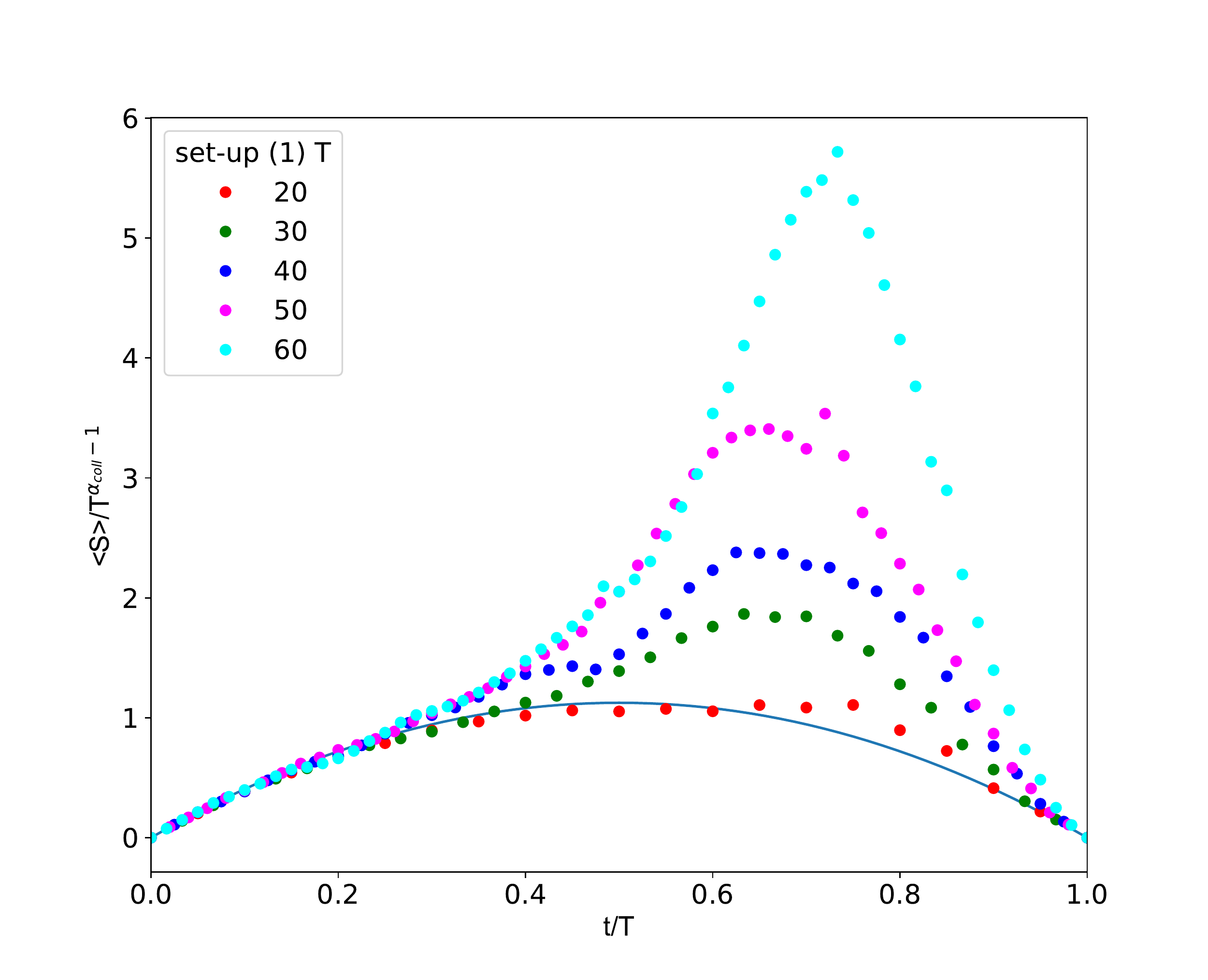}
\includegraphics[width=8.0cm, height=8cm]{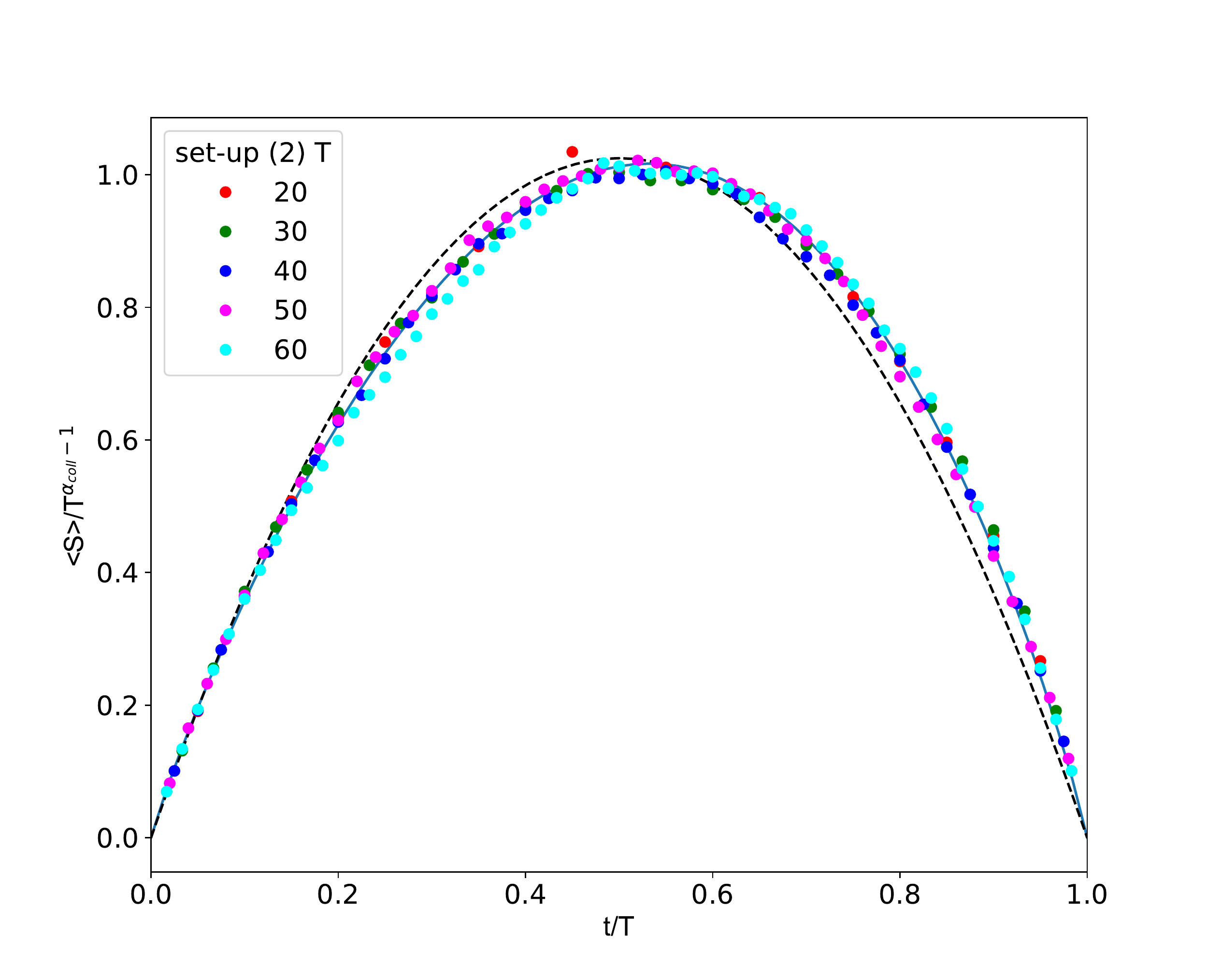}
\end{tabular}
\begin{tabular}cc
\includegraphics[width=8.0cm, height=8cm]{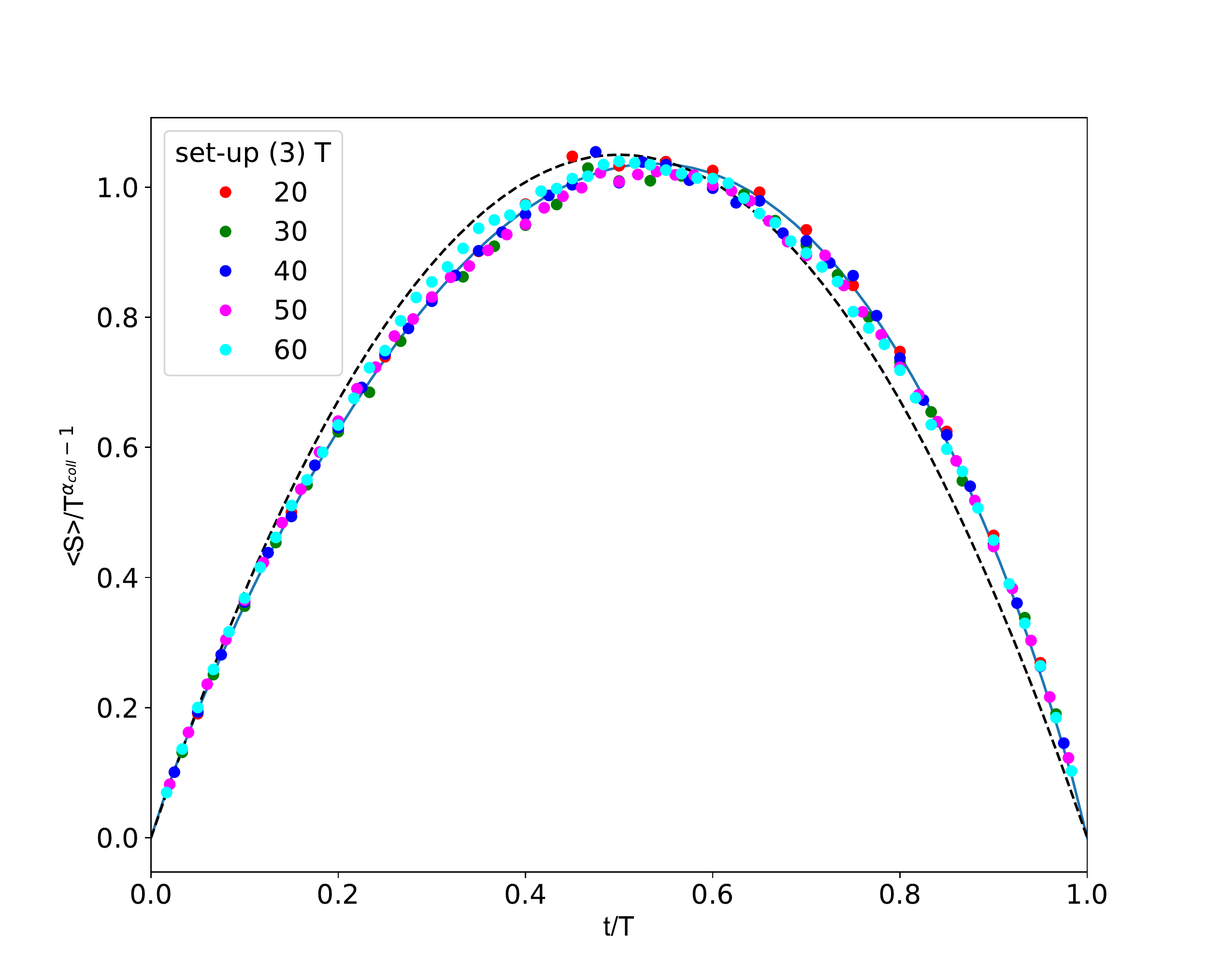}
\includegraphics[width=8.0cm, height=8cm]{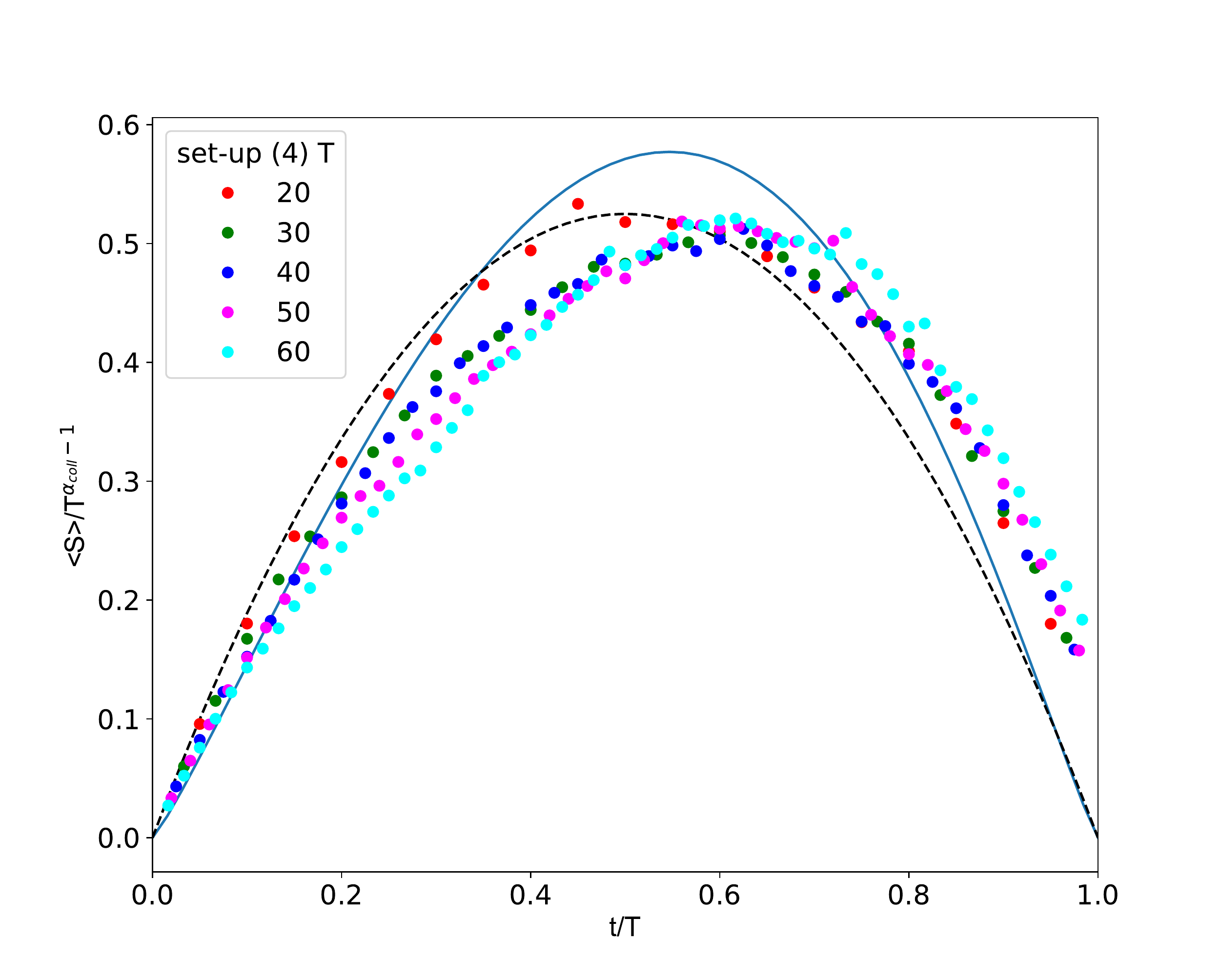}
\end{tabular}
\caption{Normalized average avalanche size $<S(t,T)>/T^{\alpha_{coll}-1}$ as a function of $t/T$ for setup 1 (upper left panel), setup 2 (upper right panel), setup 3 (lower left panel), and setup 4 (lower right panel). The dashed lines in the middle and right panels are the best fit with a parabola. The solid line is the best fit with the asymmetric function in eq. 19}
\label{avapar}
\end{figure*}

Another test to assess whether a system is close to SOC is the analysis of the shape of the avalanche profiles (e.g., Nandi et al., 2022 and references therein). At criticality, the avalanche evolution with time should follow a universal shape (a parabola for Barkhausen noise) for all durations. More precisely, the shapes corresponding to different avalanche durations should collapse on top of each other when the average size at each given time $t$ is normalized by $T^{\alpha_{coll}-1}$. To visualize the situation, we show in Figure \ref{polar_figs} three avalanches for each setup. The left panels show an avalanche that did not reach the innermost radius, the central panel shows a complete avalanche, and the right panel shows the (complete) avalanche with the largest size for that setup. The avalanche with the largest size for setup 1 fully covers the innermost radius, that is, it does not decrease in size during the evolution, but continues to expand down to the innermost radius.  The exponent $\alpha_{coll}$ is an independent estimate of the critical avalanche exponents and corresponds to the best collapse, that is, the collapse with the least scatter between the points for different avalanche durations. Figure \ref{avapar} shows $<S(t,T)>/T^{\alpha_{coll}-1}$ for avalanches with durations of $T=20$, $T=30$, $T=40$, $T=50$, and $T=60$ as a function of $t/T$. In setup 1 (upper left panel), the avalanche evolution does not follow a universal shape, but exhibits a pronounced increase in size toward the end of the avalanche for events with a long duration. In setups 2 and 3, the avalanche profile approaches a parabolic shape (Figure \ref{avapar}, upper right and lower left panels). In setup 2, $\alpha_{coll}$ is similar to $\alpha$, but the slope of the $<S>(T)$ correlation is not constant, but changes from 1.49 at low $T$ to 1.82 at high $T$.  In setup 3, all three critical exponents $\alpha$, $\alpha_1$, and $\alpha_{coll}$ are consistent with each other. It is remarkable that simple recipes mimicking the conservation of angular momentum in real disk-wind systems and feedback between the nuclear wind and accretion flow are able to robustly drive the system toward an SOC state. In setup 4, the avalanche profile is more regular than in setup 1 and closer to a parabolic shape, confirming that the inclusion of a threshold for gas accretion is enough to regularize the flow. The scatter of the points is much higher than in setups 2 and 3, however, and the avalanche profile does not fully collapse onto a universal function, indicating that the system still has not reached an SOC state. We wish to stress that the SOC state does not require the symmetry of the avalanche profile, but just the collapse onto a universal function.

The avalanche profiles in setups 2, 3, and 4 approach a parabolic shape (Figure \ref{avapar}), with a slight rightward asymmetry. This asymmetry in the literature has been detected in random processes with short and long temporal correlations in the processes (Baldassarri et al., 2003; Colaiori et al., 2004). Following Nandi et al. (2022), to quantify the level of asymmetry, we fit the average avalanche shapes in setups 2 and 3 with the asymmetric function

\begin{equation}
<S>/T^{\alpha_{coll}-1}=[1-a(t/T-0.5)][t/T(1-t/T)]^{\alpha_{coll}-1}
\end{equation},
\noindent
where $a$ is the measure of asymmetry. We found very good fits in both cases, with a best fit $a=0.24$ and $0.28$ for setups 2 and 3, respectively. These values are similar to those reported in Nandi et al. (2022) for a completely different system (a neuronal network).  A higher value ($a\sim 0.5$) is obtained for setup 4. 

\begin{figure*}
\begin{tabular}{ccc}
\includegraphics[width=5.9cm]{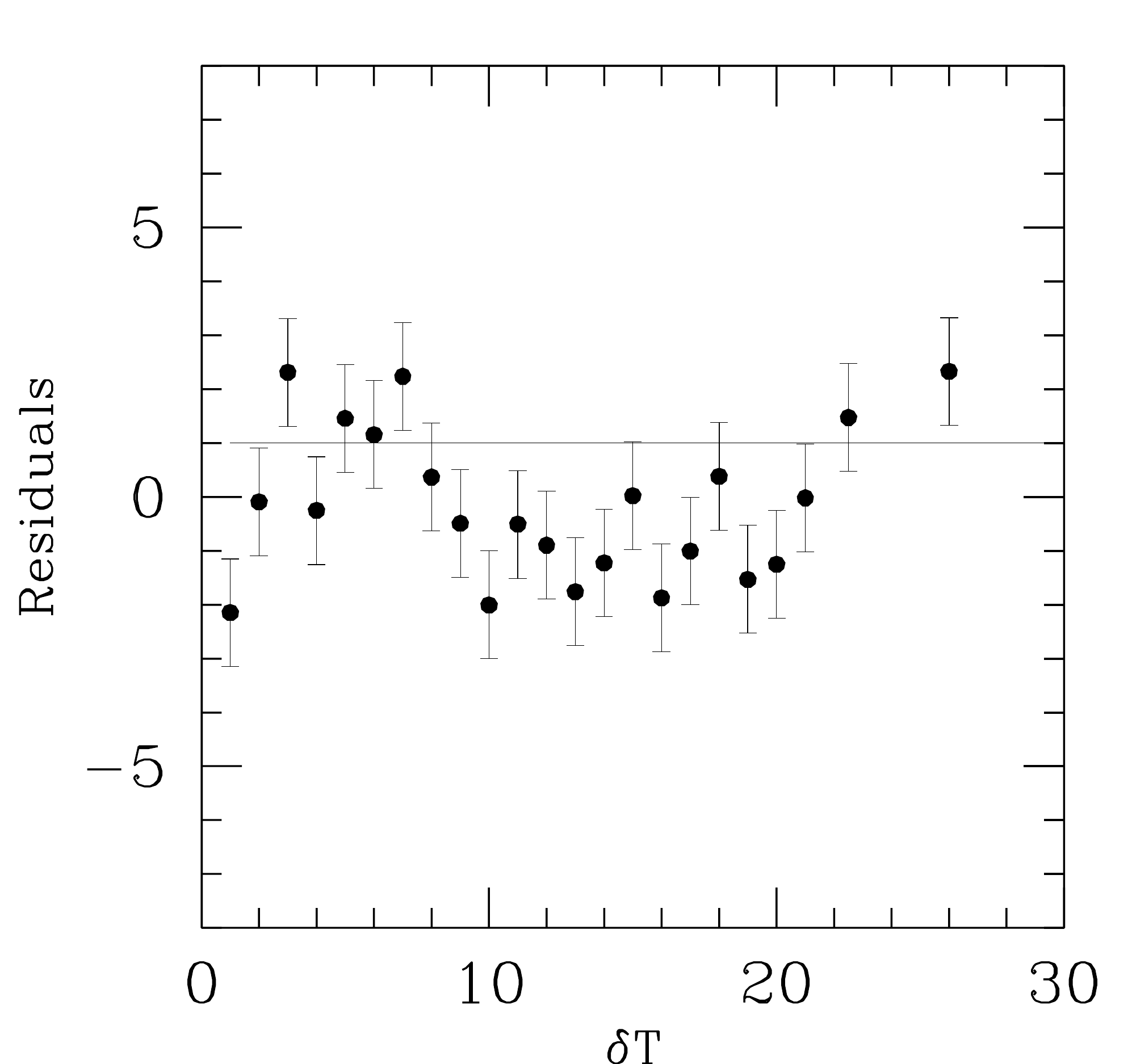}
\includegraphics[width=5.9cm]{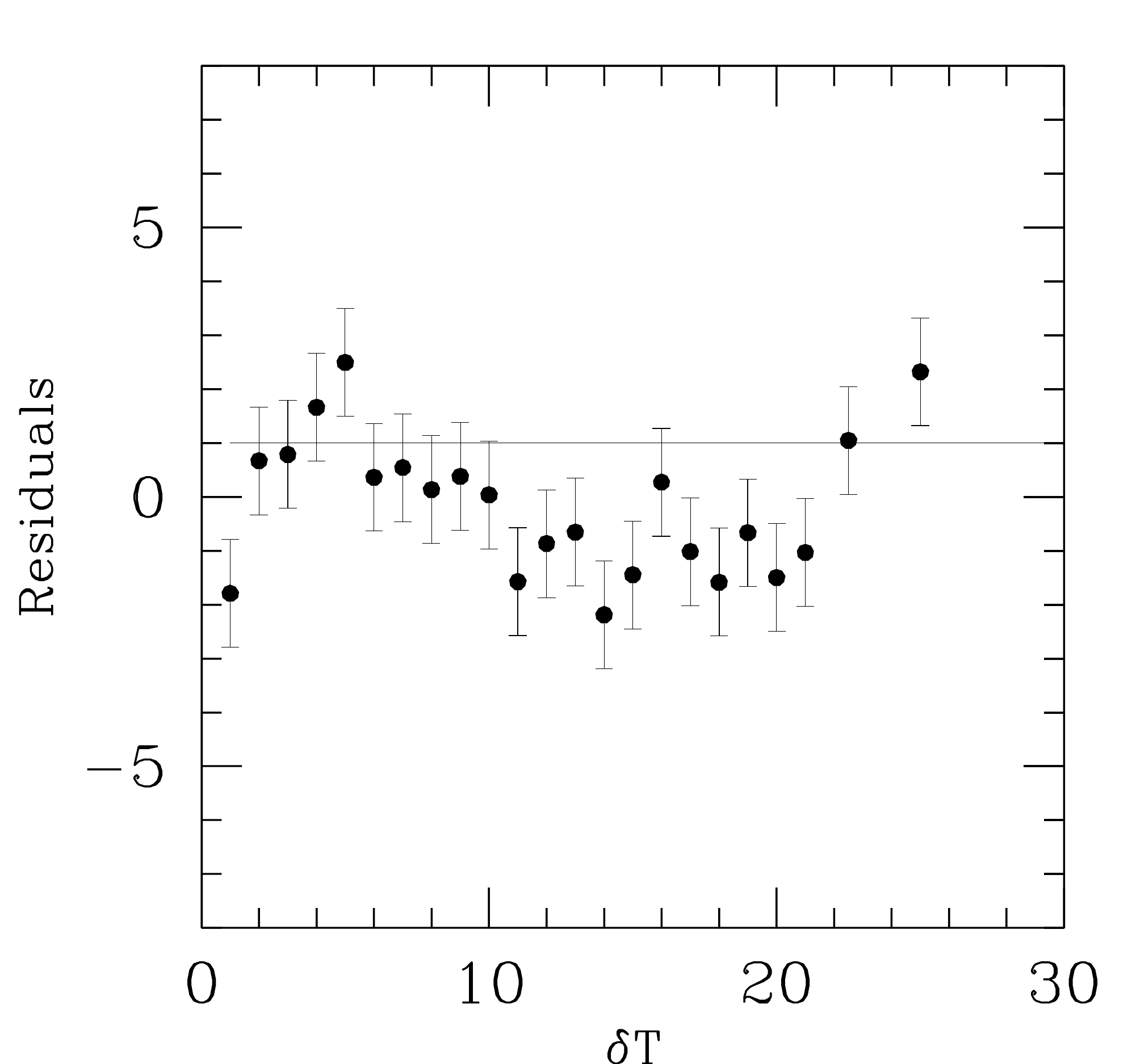}
\includegraphics[width=5.9cm]{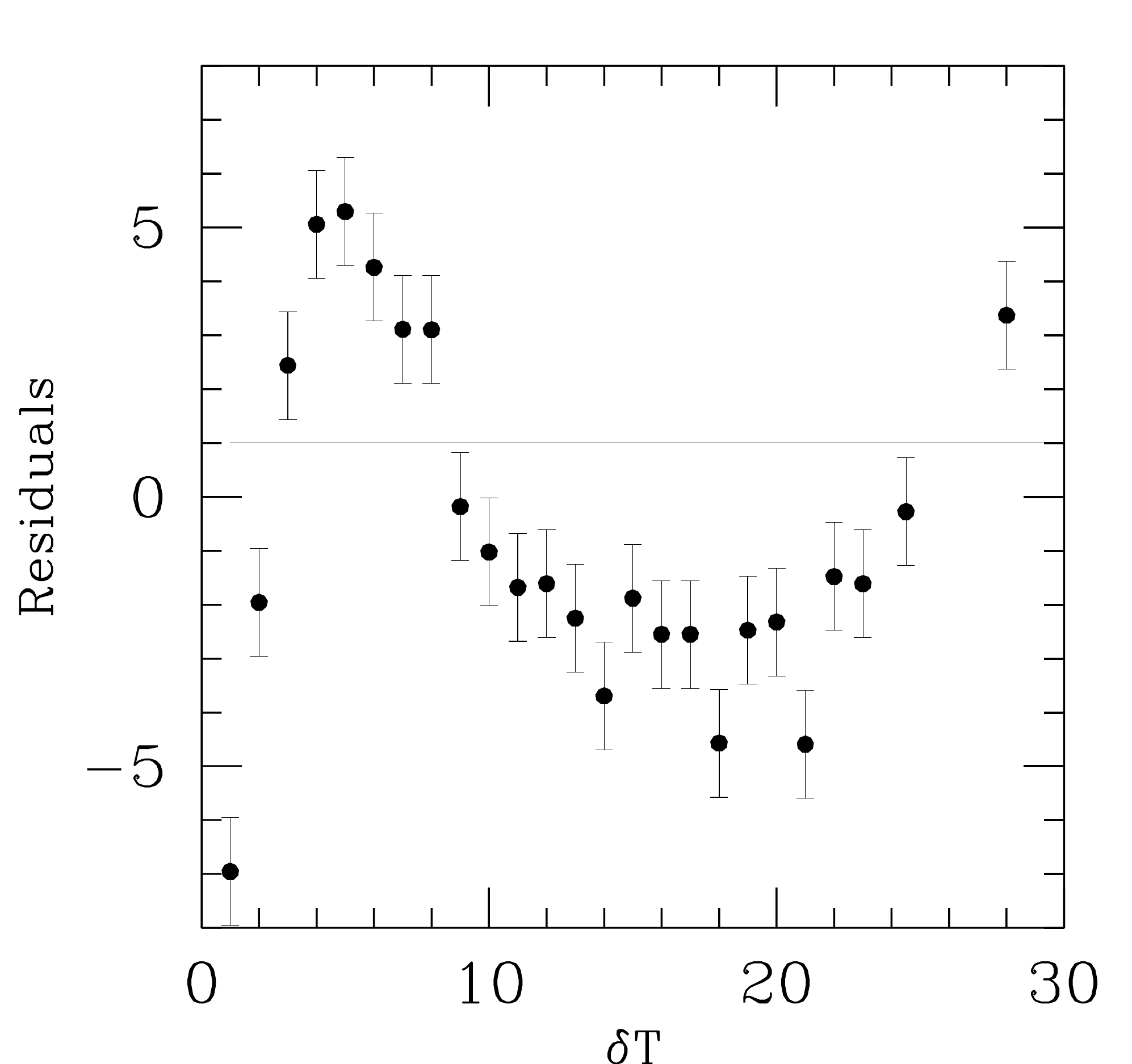}
\end{tabular}
\caption{Residuals after subtracting the best-fit exponential model from the cumulative distribution function of the waiting times $\delta t$ between one avalanche and the next. The errors are calculated using Poisson statistics. From left to right, we show setups 1, 2, and 3. The residuals for setup 3 are quite large, indicating that in this case, the distribution function of $\delta t$ is more complex than an exponential law. This indicates a correlated process.}
\label{res_dt}
\end{figure*}

To further investigate the role of feedback in bringing the system toward an SOC state, we analyzed the distributions of the waiting times between one avalanche and the following one $\delta t$. The shape of this distribution can distinguish between simple Poisson processes (as in setup 1) and correlated processes (as in setup 3), as discussed in de Arcangelis et al. (2006) and references therein. For pure Poisson processes, the distribution of $\delta t$ follows an exponential law, while for correlated processes, more complex forms can be obtained, including power laws, power laws with exponential tails, and even nonmonotonic functions. Figure \ref{res_dt} shows the residuals after subtracting from the $\delta t$ distributions of setups 1, 2, and 3 their best-fit exponential function. The distribution of the waiting times of setup 4 is identical to setup 1. While the $\delta t$ distributions obtained from setups 1 and 2 are consistent with an exponential form and display roughly regular residuals, the $\delta t$ distribution from setup 3 clearly has a more complex shape, confirming that feedback can drive the system toward an SOC state, where temporal correlation are present, even with such a minimal modeling approach. 

\section{Discussion}

The emergence of power laws, the intrinsic nonlinearity of the system, and the presence of feedback at several scales suggest that disk-wind systems could be further investigated as complex dynamical systems. We therefore developed a cellular automaton implementing the main characteristics of a disk-wind system. We studied four setups. Setup (1) is similar to the original cellular automaton introduced by Mineshighe et al. 1994, 1999. When analyzed with statistical tools inspired by critical phenomena, it is found to be supercritical, that is, with an excess of large avalanches, flattening the $f(S)$ distribution (Figure \ref{ST1}) and producing an average avalanche profile far from the expected universal scenario (Figure \ref{avapar} left panel). The reason is that avalanches cannot exit efficiently from the system in our geometry and that feedback is lacking. The only side from which an avalanche can exit the system is the innermost radius of the disk.

We introduced in setup 2 a prescription to calculate the mass that leaves the disk in a wind, inspired by MHDW systems. This reduced the accumulation of mass in each given cell, and thus reduced the avalanches size and duration, steepening the $f(S)$ and $f(T)$ distributions. The system approached an SOC state, in which avalanches can hardly extend over the full size of the system. The average avalanche profile now approached a parabolic shape (Figure \ref{avapar}, central panel). However, the scaling between $<S>$ and $T$ is not a pure power law, and the critical exponents $\alpha$ and $\alpha_1$ in Table \ref{tab:ST} are not consistent with $\alpha_{Coll}$. 

In setup (3), we further introduced feedback between the outflowing wind and the mass that accreted at the outermost radius. This further reduced the inflow rate and thus the accumulation of mass in each cell. A lower mass accumulation implies that the mass of a given cell may not be brought to a value greater than $M_{crit}$, so that this cell can give rise to accretion at a later time, thus restarting some avalanche after it has stopped. This increased the duration of these avalanches with respect to setup (2). The net effect was that this form of feedback brought the system toward an SOC state. The scaling between $<S>$ and $T$ is now a pure power law, and the critical exponents $\alpha$, $\alpha_1$ and $\alpha_{Coll}$ in Table \ref{tab:ST} are now more consistent with each other. 

We introduced in setup 4 a prescription to account for the mass that left the disk in a wind similar to the Eddington luminosity threshold and thus inspired by RDW. This was again able to regularize the flow, steepening the $f(S)$ and $f(T)$ distributions. The system is closer to SOC than that in setup 1, but worse than in setup 2 and 3 (Figure \ref{avapar}).

Some type of feeding regulated by kinetic feedback (whether magnetic or radiatively driven seems less relevant) is truly relevant in this system in general. The more intermittent and fractal-like the process (e.g., CCA), the closer it approaches SOC characteristics.

Our analysis has two main limitations. The first limitation is the rather qualitative framework in which observations are interpreted. A better approach requires the comparison of the observations to a robust physical model. We plan to extend  the WINE model (Luminari et al., 2018; 2020; 2021; 2023; Laurenti et al., 2021) by including the three main assumptions discussed above and a full MHD structure. The second limitation is that the sample of AGNs with a UFO detection is naturally biased toward detections. These limitations may be circumvented by using the well-defined and unbiased \textrm{SUBWAYS} sample of 24 AGN at z=0.1-0.5.  XRISM\footnote{\url{https://heasarc.gsfc.nasa.gov/docs/xrism}} (Kitayama et al. 2014), and on a longer timescale, Athena\footnote{\url{https://www.the-athena-x-ray-observatory.eu}} (Nandra et al. 2013, Ettori et al. 2013) high-resolution microcalorimeter observations will revolutionize our understanding of the inner AGN scale with respect to currently available data, which mostly offer only CCD resolution. Studying the distribution of $\bar \omega$ and $\lambda_w$ including upper limits will allow us to estimate the bias in using samples of published UFOs and correct for it, which will allow us to broaden the sample by including both \textrm{SUBWAYS} and previously published UFOs. This will reduce the statistical uncertainty. 

\begin{figure*}[ht]
\includegraphics[width=18cm]{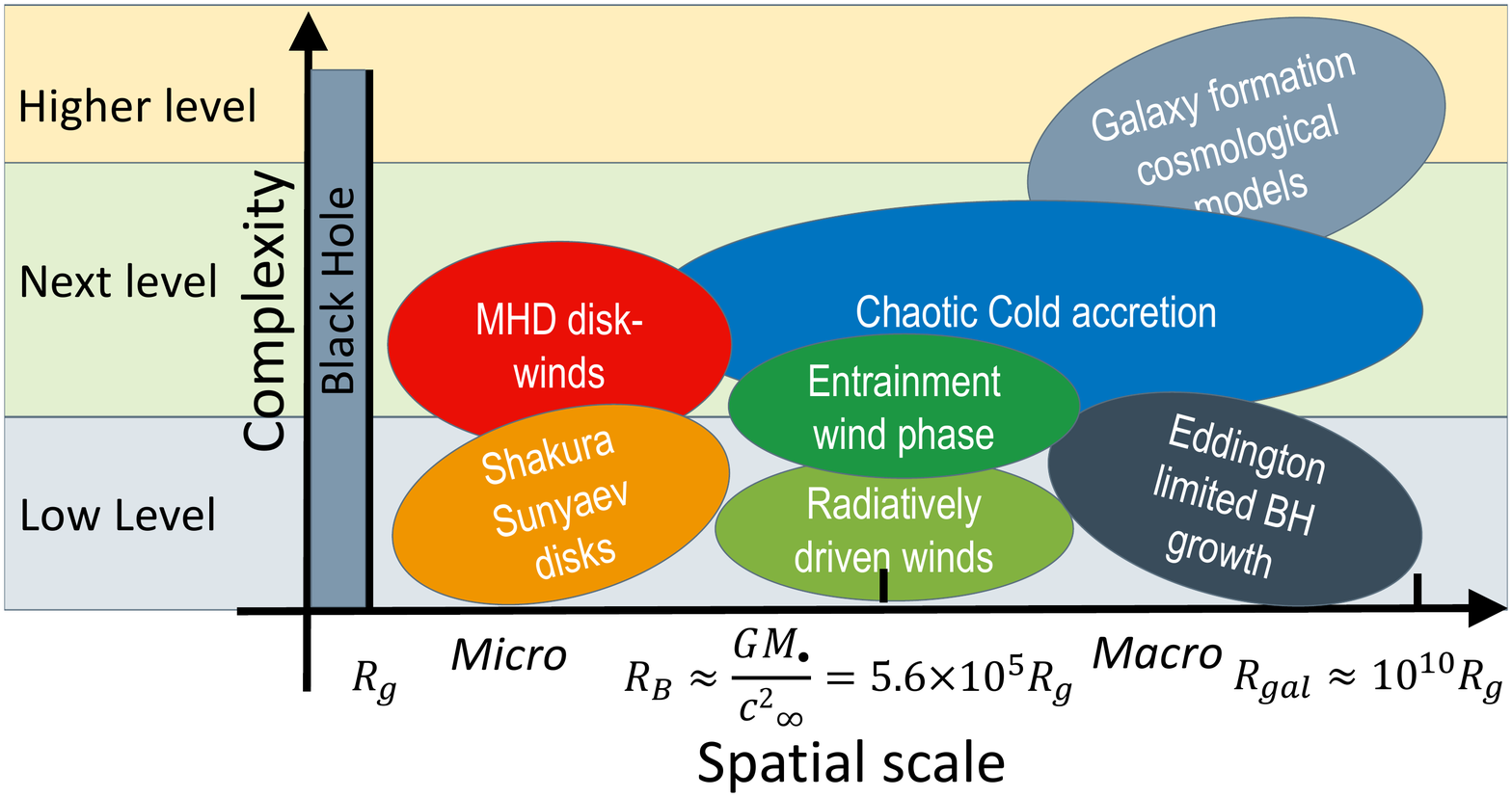}
\vspace{-2cm}
\caption{Complexity vs.~spatial scale plane for several disk-wind models and BH-galaxy coevolution models.}
\label{ds}
\end{figure*}

This is the first of a series of papers that studies the broad problem of AGN-galaxy coevolution from a different perspective, the perspective of complex dynamical systems. Figure \ref{ds} inserts popular BH disk-wind models and AGN-galaxy coevolution models into a  {\it complexity - spatial scale} plane. Following GS17 (updated in Gaspari et al., 2020), we divided the spatial scale into three broad regions, the microscale, spanning roughly from the BH event horizon to well within the BH sphere of influence radius, the mesoscale, around the BH sphere of influence from $10^3 r_S$ to $10^6 r_S$, and the macroscale, outward up to the edge of the Galaxy and its circumgalactic medium. This paper is dedicated to disk-wind systems and therefore studies microscales. On scales encompassing the BH sphere of influence, between the outer accretion disk and the nucleus of the galaxy (i.e., 100-100.000 $r_S$, or between a fraction of a parsec to tens of hundred parsecs) several wind models have been discussed, the RDWs, momentum- and energy-driven winds (e.g., Faucher-Giguere \& Quataert 2012; Zubovas \& King 2012), and the entrainment wind phase of GS17. 

On macroscales, three main broad scenarios have been discussed in the literature at length. The first scenario, the Eddington-limited BH growth, occupies the lower right part of the diagram in Figure \ref{ds}. This scenario dates back to the work of Silk \& Rees (1998), Fabian (1999), King (2003), and Granato et al. (2004; see the reviews of Fabian 2012; King \& Pounds 2015 and Marsden et al., 2020). Briefly, BH growth can only proceed when AGN winds are not strong enough to displace the accreting gas (negative feedback).  On one hand, the wind thrust is proportional to $L_{AGN}/c$, and $L_{AGN}\leq L_{Edd}$, and on the other hand, the weight of the accreting gas is proportional to $\sigma^4$ (where $\sigma$ is the bulge velocity dispersion), assuming an isothermal gas distribution. Thus, for a given $\sigma$, the BH can only grow up to a critical mass, giving rise to the $M_{BH} - \sigma$ relation or to a ridge in the distribution of $M_{BH}$ for each $\sigma$, the BH adjustment scenario of Volonteri (2012), and the $M_{BH}$-X-ray scaling relations (Gaspari et al. 2019), which have been observed to be even tighter than the above optical relations (e.g., see the $M_{BH}$ -- X-ray temperature relation). This broad scenario introduces the crucial concept of BH wind feedback that limits the growth of the same BH, but this simplified view has little room for random processes. 

The second promising scenario, which we will also study in more depth with a follow-up study, is the CCA introduced by Gaspari et al. (2013; 2017; 2019) and later by Voit et al. (2015; 2018). Briefly, gas in the BCG halo cools down and condenses into cold clouds and filaments. Random collisions between clouds and filaments can efficiently dissipate their angular momentum, thus enabling efficient inward flows toward the galaxy nucleus and its supermassive BH, starting both star formation in filaments close to the BCG central regions and an AGN cycle. The triggered AGN feedback then transports gas mass and energy back to the macroscale, stopping the star formation, reheating the bulk of the ICM to a higher entropy level, virtually ending also the process, regardless of complexity, that caused the mass deposition rate, and thus completing a single AGN cycle.  Thereafter, the gas in the halo, no longer heated, can again condense, giving rise to a new star formation - AGN cycle. This is reflected in the correlations between nuclear activity, star formation rate, and entropy of the ICM in cluster cores, but with large scatter. Correlations of this type have been found: the power to excavate cavities in the ICM is proportional to the X-ray luminosity of the cool core and to the AGN radio luminosity. Similarly, the tight correlations between the SMBH mass and X-ray properties (Gaspari et al. 2019) seem to arise from the CCA raining cycles integrated over several billion years. Only the galaxies in clusters/groups with a low inner entropy (short cooling time) have an active SMBH and are actively forming stars (Cavagnolo et al. 2008, 2009). In this model, BH feedback is again used to regulate the BH growth, and random processes are introduced to both modulate gas accretion toward the galaxy nucleus and its feedback to the gas cooling. In these models, random collisions create a self-similar time variability with a flicker-noise 1/f power spectrum (Gaspari et al. 2017), similar to SOC systems. CCA is thus an excellent example of a chaotic and emergent process in astrophysics, in which the collective behavior of a large number of microscopic constituents (multiphase clouds) drives a completely novel macroscopic physical process (flickering BH accretion and rain). 

Other forms of clumpy accretion have been proposed, such as the chaotic accretion through a sequence of randomly oriented merger episodes (King \& Pringle 2006; King et al. 2008). Ishibashi \& Courvoisier (2009, 2012) also suggested that shocks between accreting clouds can result in X-ray characteristic timescales/spectra with a composition of flicker and red-noise power laws. In the future, it will be interesting to compare the different flavors of chaotic/clumpy accretion models using the tools adopted in Section 5.

The third broad scenario includes cosmological simulations using both semi-analytic models (SAM; Fontanot et al., 2020; Menci et al., 2014; 2023 and references therein) and hydrodynamic numerical models (e.g., Di Matteo et al., 2005; Costa et al., 2014; Sijacki et al., 2015). These simulations follow the growth of density perturbations of the dark matter field and incorporate baryon physics at some level of detail to describe star formation, galaxy and BH growth, and stellar and BH feedback processes. These models by construction include random processes and also include feedback on both galaxy and BH growth from star formation and BH winds and jets. The nuclear accretion in these models is driven by galaxy interactions and by internal galaxy instabilities, such as bars and clumps. These are again random processes. At the heart of star formation and nuclear BH accretion processes are instabilities in the gas disks that can grow and produce strong inward gas flows. Disk instabilities and gravitational disk fragmentation are suppressed or reduced when the fraction of gas to total mass is low (e.g., Lilly et al. 2013). This may occur because AGN feedback removes cold gas from the disk, but other mechanism can be at work as well, such as supernova feedback (especially in low-mass systems), inefficient gas accretion above a threshold galaxy mass (Dekel \& Birnboim 2006), and mechanisms impeding inward gas transport (Genzel et al. 2014).

We plan to rethink these scenarios in an attempt to interpret micro- and macroscale observations in the framework of complex dynamical systems by searching for evidence (or absence of evidence) of nonlinearity, universality, and the development of scaling laws, self-organization, criticality, and/or other complex behaviors on scales from the accretion disk in the BH sphere of influence, the bulge, the galaxy disk, and the circumgalactic medium. Finding evidence of the same complex dynamical behavior and universality class (similar critical exponents), triggered by BHs on systems from the micro- to the macroscale would highlight the existence of unified principles and would strongly point towards the BH necessity scenario. 

\section{Conclusions}

We have compiled a large sample of previously published UFO observations and interpreted them in the framework of a scenario including three main assumptions: 1) links between accretion and wind emission as in MHDW and RDW, that is, $\dot M_{accr}=(1/\bar \omega) \dot M_{\rm w}$ in the former case, or a threshold in disk luminosity above which the accretion is blocked in the latter case; 2) an Eddington-like limit on the mass-accretion rate due to the wind thrust that contrasts the accretion flow at the Bondi radius; and 3) a random distribution of accretion/wind events. We find that the distribution functions of $\bar \omega=\dot M_{\rm w}/\dot M_{accr}$ and $\lambda_w=\dot M_{\rm w}/\dot M_{Edd}$ can be described by power laws above some thresholds, suggesting scale-invariance and universal emergent properties in disk-wind systems. Under this assumption, the wind activity timescale is short, much shorter than the AGN timescale, which agrees with other determinations (e.g., Nardini \& Zubovas 2018; King \& Pounds 2015), and that it is proportional to $\sqrt {\bar \omega}$. When the population properties ($\bar \omega$ and $\lambda_w=\dot M_{\rm w}/\dot M_{Edd}$) are drawn from independent snapshots, it might be argued that the wind activity history in each single source can be statistically described by the population functions. If this is the case, the emerging scenario is that of many relatively small wind events for each main wind event in each AGN activity cycle. Because $\lambda=\dot M_{BH}/\dot M_{Edd}=\lambda_w/\bar \omega$,  $\lambda$ should also follow a power-law distributions down to an inflection point. The BH growth obtained during each AGN activity cycle is the sum of the growth at each $\lambda$, weighted for the time spent at that $\lambda$. When this time is similar to the wind activity timescale, then $\lambda \Delta t \propto \lambda_w/\sqrt{\bar \Omega}$. The slope of the $\lambda \Delta t$ distribution is similar to that of the $\lambda$ distributions estimated by Bongiorno et al (2012) and Aird et al. (2012, 2018) for X-ray selected AGN samples, but it is somewhat flatter than that found by Ananna et al. (2022) for a sample of local AGN. 

Furthermore, we developed a simple theoretical/numerical model based on a cellular automaton with the goal of testing whether by using simple feedback rules, it is possible to build a dynamical system based on BH physics that approaches the state of SOC, beyond the purely Poisson baseline. In all our simulations, feedback seems to be a necessity to drive the system toward an SOC state. This is encouraging because current models of SMBH feeding, such as CCA, show that fractal distributions and flickering variability are crucial for achieving a consistent self-regulation (Gaspari et al. 2020, for a review). 

The main question we started with was whether BH feedback is a coincidence or a necessity for the evolution of galaxies, groups, and clusters of galaxies. Our results consolidate the view that feedback is a necessity for the disk-wind system. Finding similar universality class and statistical properties (scale-invariance or feedback-driven SOC) on large scales as well would point toward the BH necessity scenario. 
\newline

\noindent
{\bf Acknowledgements}\\
FF acknowledge support from PRIN MUR 2017 "Black hole winds and the baryon life cycle of galaxies: the stone-guest at the galaxy evolution supper". FF and AL acknowledge support from the European Union Horizon 2020 Research and Innovation Framework Programme under grant agreement AHEAD2020 n. 871158. MG acknowledges partial support by HST GO-15890.020/023-A, the "BlackHoleWeather" program, and NASA HEC Pleiades (SMD-1726). PT acknowledges financial support through grant PRIN MUR 2017WSCC32: “Zooming into dark matter and proto-galaxies with massive lensing clusters”.  
FF thanks N. Menci and E. Piconcelli for initial discussions on the UFO sample and wind trust feedback. We acknowledge very useful discussion during the workshop Dynamical complexity in astrophysical contexts: https://indico.ict.inaf.it/event/2307/ 
FF dedicates this paper to the memory of Prof. Giuseppe Cesare Perola, for his vision and constantly present guidance.

\newpage
\begin{appendix}
    
\section{UFO system sample}

The UFO systems used in this paper are presented in Table \ref{tab:app}. The table includes AGN observed by XMM-Newton (most of them), Chandra, and Suzaku. The sample only includes positive detections and is therefore neither complete nor unbiased. We accounted for this bias by calculating a detection probability function based on a relatively large sample of XMM-Newton observations of local AGNs. We used this probability function to correct for the observed outflow rate distribution functions, as discussed in the next section.  

\begin{table*}[!ht]
\label{tab:app}
\caption{Table A1: UFO system sample}
\begin{tabular}{lcccccccccl} 
\hline
Name$^a$ & redshift & $\log L_{\rm bol}$ & $v_{\rm max}$ & $\log M_{\rm BH}$ &  $\log R_{\rm min}$ & $\log R_{\rm ion}$ & $\log \dot M_{\rm min}$ & $\log \dot M_{\rm ion}$ & $\log \dot M$ & References \\
 &  & erg s$^{-1}$ &  v/c & M$_\odot$  & $r/r_S$ & $r/r_S$  & M$_\odot$/yr & M$_\odot$/yr & M$_\odot$/yr  & \\
\hline
NGC4151 &   0.0033 & 43.9 & 0.106 &  7.1 &    89 &   322 & -2.66 & -2.10 & -2.26 & 1,2 \\ 
IC4329A &   0.0161 & 45.1 & 0.098 &  8.1 &   104 &    61 & -1.64 & -1.86 & -1.86 & 1 \\ 
Mrk509 &   0.0344 & 45.2 & 0.173 &  8.2 &    33 &    42 & -1.33 & -1.24 & -1.24 & 1 \\ 
Mrk509 &   0.0344 & 45.4 & 0.138 &  8.2 &    53 &    72 & -1.23 & -1.10 & -1.10 & 1 \\ 
Mrk509 &   0.0344 & 45.4 & 0.217 &  8.2 &    21 &  2680 & -2.23 & -0.13 & -1.83 & 1 \\ 
Ark120 &   0.0327 & 45.5 & 0.287 &  8.2 &    12 &  1615 & -2.33 & -0.21 & -1.94 & 1 \\ 
Mrk79 &   0.0809 & 44.9 & 0.092 &  7.9 &   118 &    85 & -0.80 & -0.94 & -0.94 & 1,3 \\ 
NGC4051 &   0.0023 & 43.3 & 0.037 &  6.3 &   730 &  1092 & -3.29 & -3.12 & -3.12 & 1 \\ 
NGC4051 &   0.0023 & 43.0 & 0.202 &  6.3 &    25 &  2229 & -3.33 & -1.37 & -2.93 & 1 \\ 
Mrk766 &   0.0129 & 44.4 & 0.088 &  6.8 &   129 &   547 & -2.18 & -1.55 & -1.78 & 1,4 \\ 
Mrk841 &   0.0364 & 44.9 & 0.055 &  8.5 &   331 &   809 & -1.27 & -0.89 & -0.89 & 1,5 \\ 
1H0419 &   0.1040 & 45.6 & 0.079 &  8.1 &   160 &   973 & -0.59 &  0.19 & -0.20 & 1,6 \\ 
Mrk290 &   0.0296 & 44.6 & 0.163 &  7.2 &    38 &   315 & -1.64 & -0.71 & -1.24 & 1,3 \\ 
Mrk205 &   0.0708 & 45.2 & 0.100 &  7.9 &   100 &    47 & -1.00 & -1.34 & -1.34 & 1,7 \\ 
MCG-5-23 &   0.0085 & 44.5 & 0.116 &  7.4 &    74 &   379 & -2.09 & -1.38 & -1.69 & 1,8 \\ 
NGC4507 &   0.0118 & 44.4 & 0.199 &  6.4 &    25 &  8094 & -3.95 & -1.45 & -3.56 & 1 \\ 
NGC7582 &   0.0052 & 43.3 & 0.285 &  7.1 &    12 &    76 & -2.04 & -1.25 & -1.64 & 9 \\ 
3C111 &   0.0485 & 45.7 & 0.072 &  8.1 &   193 &   387 & -0.81 & -0.51 & -0.51 &  10 \\ 
3C390.3 &   0.0561 & 45.5 & 0.145 &  8.5 &    48 &     3 & -0.06 & -1.21 & -1.21 & 10,2 \\ 
ESO103 &   0.0132 & 44.6 & 0.056 &  7.4 &   319 &  3786 & -2.40 & -1.33 & -2.01 & 10 \\ 
MR2251 &   0.0640 & 46.0 & 0.137 &  8.7 &    53 & 47660 & -1.49 &  1.46 & -1.09 & 10\\ 
Mrk279 &   0.0344 & 44.0 & 0.220 &  7.5 &    21 &    34 & -1.51 & -1.29 & -1.29 & 10,3\\ 
Mrk766 &   0.0129 & 44.1 & 0.039 &  6.1 &   657 &  9509 & -2.84 & -1.68 & -2.45 & 10,4 \\ 
NGC4051 &   0.0023 & 43.2 & 0.018 &  6.3 &  3086 &  2178 & -2.21 & -2.36 & -2.36 & 10\\ 
NGC4151 &   0.0033 & 43.6 & 0.055 &  7.1 &   331 &  3533 & -2.69 & -1.67 & -2.30 & 10\\ 
NGC5506 &   0.0062 & 44.4 & 0.246 &  6.9 &    17 &   109 & -2.38 & -1.56 & -1.99 & 10,3 \\ 
SWJ2127 &   0.0147 & 44.5 & 0.231 &  7.2 &    19 &   951 & -2.42 & -0.71 & -2.02 & 10 \\ 
4C74.26 &   0.1040 & 46.3 & 0.185 &  9.6 &    29 &  6000 & -0.72 &  1.59 & -0.32 & 10,11 \\
NGC2992 &   0.0077 & 44.4 & 0.215 &  7.7 &    22 &     9 & -1.32 & -1.70 & -1.70 & 12\\ 
NGC2992 &   0.0077 & 44.4 & 0.305 &  7.7 &    11 &     6 & -1.47 & -1.75 & -1.75 & 12 \\ 
MCG-03-58 &   0.0315 & 45.5 & 0.075 &  8.0 &   178 &     3 & -0.04 & -1.81 & -1.81 & 13 \\ 
MCG-03-58 &   0.0315 & 45.5 & 0.200 &  8.0 &    25 &     4 & -0.62 & -1.38 & -1.38 & 13\\ 
Mark231 &   0.0422 & 45.7 & 0.067 &  8.4 &   223 &   600 & -0.07 &  0.36 &  0.33 & 14 \\ 
IrasF051 &   0.0426 & 45.7 & 0.110 &  8.3 &    83 &    42 & -0.37 & -0.66 & -0.66 & 15 \\ 
I17020 &   0.0604 & 44.7 & 0.091 &  6.8 &   121 &   125 & -1.25 & -1.24 & -1.24 & 16 \\ 
IZwicky1 &   0.0611 & 45.5 & 0.265 &  7.5 &    14 &    49 & -1.16 & -0.62 & -0.76 & 17 \\ 
PG1448 &   0.0645 & 45.3 & 0.150 &  7.3 &    44 &    26 & -1.46 & -1.70 & -1.70 & 18 \\ 
I13224 &   0.0658 & 44.5 & 0.236 &  6.3 &    18 &  1615 & -2.81 & -0.85 & -2.41 & 19 \\ 
I00521 &   0.0689 & 45.8 & 0.405 &  7.7 &     6 &  2334 & -2.51 &  0.07 & -2.11 & 20 \\ 
PG1211 &   0.0809 & 45.6 & 0.110 &  8.0 &    83 &  6283 & -1.23 &  0.66 & -0.83 & 21 \\ 
PDS456 &   0.1840 & 47.1 & 0.250 &  9.2 &    16 &     3 &  0.63 & -0.08 & -0.08 & 22 \\ 
PDS456 &   0.1840 & 47.1 & 0.450 &  9.0 &     5 &    44 & -0.28 &  0.68 &  0.12 & 22 \\ 
I11119 &   0.1890 & 46.2 & 0.255 &  8.3 &    15 &    26 &  0.64 &  0.86 &  0.86 & 23 \\ 
TON28 &   0.3290 & 46.4 & 0.280 &  8.0 &    13 &    87 & -1.10 & -0.27 & -0.70 & 24\\ 
PID352 &   1.5900 & 46.0 & 0.140 &  8.7 &    51 &  2743 & -0.14 &  1.59 &  0.26 & 25\\ 
HS1700 &   2.7348 & 47.7 & 0.380 &  9.6 &     7 &   867 &  0.72 &  2.81 &  1.11 & 26\\ 
HS1700 &   2.7348 & 47.3 & 0.270 &  9.6 &    14 &  2178 &  0.63 &  2.84 &  1.03 & 26 \\ 
HS1700 &   2.7348 & 47.3 & 0.570 &  9.6 &     3 &  2501 &  0.26 &  3.17 &  0.66 & 26 \\ 
HS0810 &   1.5100 & 45.0 & 0.100 &  8.6 &   100 &   185 &  0.01 &  0.27 &  0.27 & 27 \\ 
HS0810 &   1.5100 & 45.1 & 0.120 &  8.6 &    69 &   144 & -0.00 &  0.31 &  0.31 & 27 \\ 
HS0810 &   1.5100 & 45.1 & 0.410 &  8.6 &     6 &   213 & -0.61 &  0.95 & -0.21 & 27 \\ 
MG0414 &   2.6400 & 46.0 & 0.280 &  9.3 &    13 &    21 &  0.72 &  0.94 &  0.94 & 28 \\ 
APM08 &   3.9110 & 47.3 & 0.160 &  9.8 &    39 &    87 &  1.54 &  1.89 &  1.89 & 29 \\ 
APM08 &   3.9110 & 47.3 & 0.350 &  9.8 &     8 &  5472 &  0.20 &  3.03 &  0.60 & 29 \\ 
\hline
\hline
\end{tabular}

1 - Tombesi et al. 2010;
2 - Peterson et al. 2004;
3 - Ricci et al. 2017;
4 - Bentz et al 2018;
5 - Vestergaard \& Peterson 2006;
6 - Tilton \& Shull 2013;
7 - Kelly \& Bechtold 2007;
8 - Caglar et al. 2020;
9 - Piconcelli et al 2007, Bianchi et al. 2009, Rivers et al. 2015;
10 - Gofford et al, 2015;
11 - Gu, Cao, Jiang 2001; 
12 - Marinucci et al. 2018;
13 - Braito et al. 2018; 
14 - Feruglio et al. 2015, Nardini et al. 2018;
15 - Smith et al. 2019, Onori et al. 2017; 
16 - Longinotti et al 2015, 2018; 
17 - Reeves \& Braito 2019;
18 - Kosec et al. 2020, Laurenti et al 2021;
19 - Parker et al. 2017, Jiang et al. 2018, Alston et al. 2020;
20 - Walton et al. 2019;
21 - Pounds \& Reeves 2009;
22 - Nardini et al. 2015, Luminari et al. 2018, Reeves et al. 2018;
23 - Tombesi et al. 2015, Nardini et al. 2018;
24 - Nardini et al. 2019;
25 - Vignali et al. 2015;
26 - Lanzuisi et al. 2012;
27 - Chartas et al. 2016;
28 - Dadina et al. 2018;
29 - Chartas et al. 2009.
\end{table*}

\section{Correction for incompleteness}

To better assess the statistical properties of the UFO sample, we accounted for selection effects and associated incompleteness of the sample. The main selection effect at work in the study of absorption features in the X-ray spectra of AGN is certainly the difficulty of detecting faint features at energies at which the effective area of X-ray telescopes decreases sharply. Statistical and systematic errors both limit our capability to identify and quantify faint UFOs. To evaluate the efficiency in detecting faint high-energy absorption lines in AGN spectra, we considered a sample of about 100 XMM observations of local AGN with exposure times between 10\,ks and 100\,ks (Tombesi et al. 2010). We then calculated the limit equivalent width ($EW$) in the 7-9 keV energy range using an analytic approach and proper spectral simulations to validate it,
\begin{equation}
    EW_{\rm lim}=\frac{S}{N}\sqrt{\frac{\Delta E(E)}{A_{eff}(E)\times T \times F(E)}},
\end{equation}
\noindent
where $\Delta E(E)$ is the energy resolution in keV, $A_{eff}(E)$ is the effective area in cm$^2$, $T$ is the exposure time in s, and $F(E)$ is the source spectrum in photon/s/cm$^{-2}$. We assumed an $S/N=4$ to evaluate the $EW_{\rm lim}$ at 7, 8, and 9 keV. We compared $EW_{\rm lim}$ to that obtained at the same energies using proper spectral simulations and found a good match. 

\begin{figure}[ht]
\includegraphics[width=8.5cm]{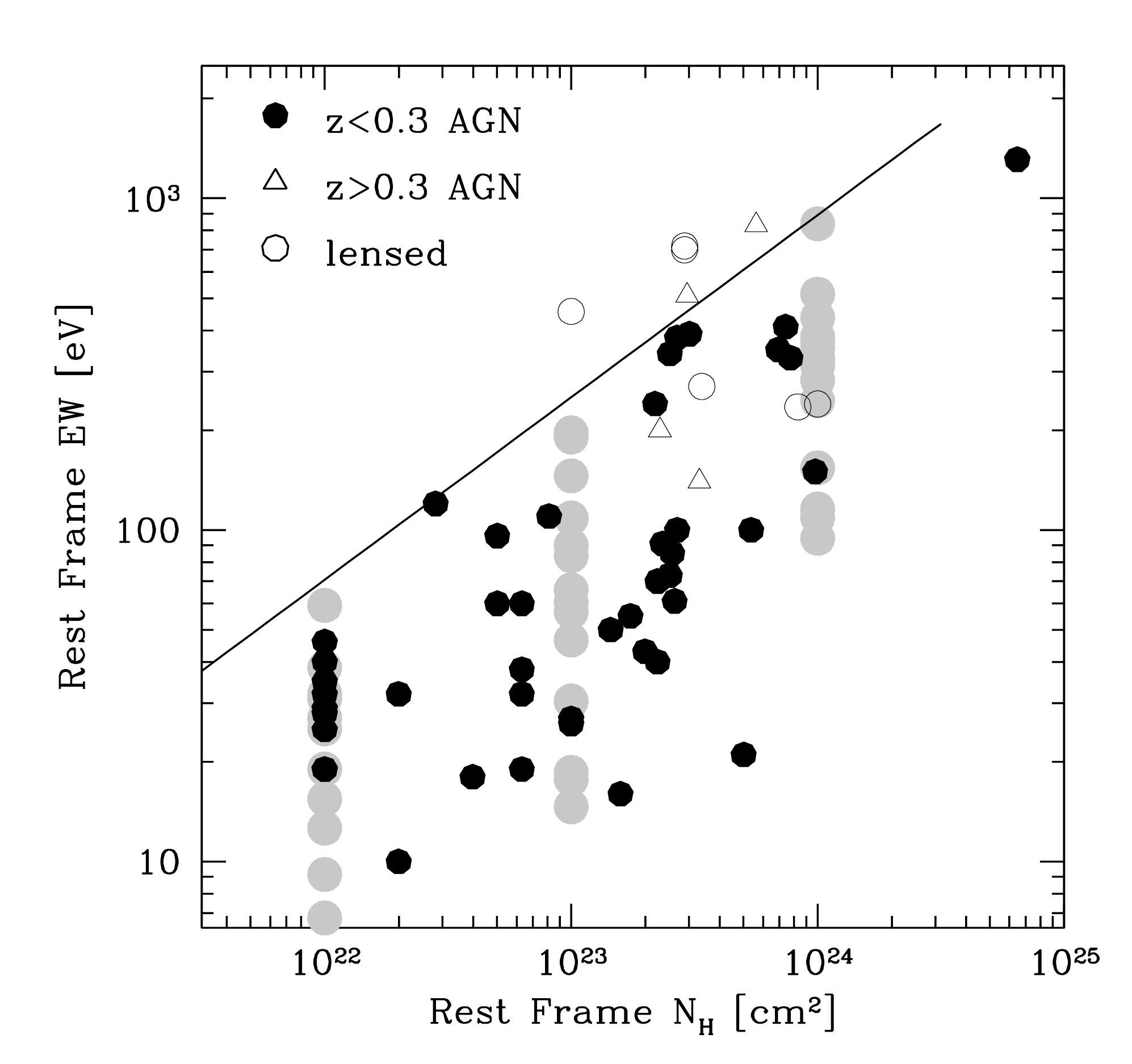}
\caption{Equivalent width $EW$ of the absorption line system between 7 and 9 keV as a function of the estimated $N_H$ for the UFOs in Table \ref{tab:app} (black points). The gray points are from a large number of XSTAR models with $10^{22}<N_H<10^{24}$ cm$^{-2}$, $10^3<\xi<10^5$ ergs/s cm, and $1000<b<30000$ km/s. The black line is the upper envelope of all XSTAR models and all observed points. It represents the minimum $N_H$ corresponding to a given $EW$.
}
\label{ewnh}
\end{figure}

To calculate a limit on the mass-outflow rate $\dot M_{\rm w}$ using eq. 2, we converted the limit equivalent width into a limit column density. The relation between $EW$ and $N_H$ is complex because it depends on the considered transition (FeXXVI Ly$\alpha$ or FeXXV He$\alpha$) on the ionization parameter $\xi$ and on the intrinsic velocity broadening $b$ (e.g., Tombesi et al. 2011). We plot in fig. \ref{ewnh} the measured $EW$ and the estimated $N_H$ for the sources of our sample. Except for of a few lensed sources and a few z$>0.3$ AGN, most points are confined to below the black line. 

We also computed the $EW$ of the absorption systems between 7 and 9 keV in a series of XSTAR models with $10^{22}<N_H<10^{24}$cm$^{-2}$, $10^3<\xi<10^5$ergs/s cm, and $1000<b<30000$km/s (gray points in fig. \ref{ewnh}). These models were also confined to the right of the black line, which can then be considered as the envelope to the $EW$ distribution as a function of $N_H$, $\xi$, and $b$. We adopted the relation $log(EW)=log(N_H)*0.55-10.5$ to convert the estimated $EW_{\rm lim}$ into minimum column densities $N_H(min)$. We finally ran a Monte Carlo simulation to associate with each $N_H({\rm min})$ a limit $\dot M_{\rm w}$ using eq. 2, where $M_{\rm BH}$ and $v_{\rm w}$ were randomly sampled from distributions resembling the observed distributions. We repeated the same procedure hundreds of times to build large samples of limit $\dot M_{\rm w}$. The normalized distribution function of the limit $\dot M_{\rm w}$ gives the probability of detecting an outflow rate higher than the given limit $\dot M_{\rm w}$. We finally used these probability functions to correct for the observed distribution function to estimate the intrinsic $\dot M_{\rm w}$ distributions. 

\end{appendix}

\end{document}